\documentclass[twocolumn]{aastex631}

\usepackage{amsmath}

\definecolor{mygreen}{rgb}{0.2,0.7,0.4}
\usepackage{xcolor}

\newcommand{\unit}[1]{\ensuremath{\,\mathrm{#1}}}

\shorttitle{EOS Dependence of GWs in CCSNe}
\shortauthors{Eggenberger Andersen et al.}

\graphicspath{{./}{figures/}}

\begin{document}

\title{Equation of State Dependence of Gravitational Waves in Core-Collapse Supernovae}

\author[0000-0002-9660-7952]{Oliver Eggenberger Andersen}
\affiliation{The Oskar Klein Centre, Department of Astronomy,\\ Stockholm University, AlbaNova, SE-106 91 Stockholm, Sweden}

\author[0000-0001-6773-7830]{Shuai Zha}
\affiliation{The Oskar Klein Centre, Department of Astronomy,\\ Stockholm University, AlbaNova, SE-106 91 Stockholm, Sweden}

\author[0000-0003-0849-7691]{Andr\'e da Silva Schneider}
\affiliation{The Oskar Klein Centre, Department of Astronomy,\\ Stockholm University, AlbaNova, SE-106 91 Stockholm, Sweden}

\author[0000-0003-2489-1808]{Aurore Betranhandy}
\affiliation{The Oskar Klein Centre, Department of Astronomy,\\ Stockholm University, AlbaNova, SE-106 91 Stockholm, Sweden}

\author[0000-0002-5080-5996]{Sean M. Couch}
\affiliation{Department of Physics and Astronomy, Michigan State University, East Lansing, MI 48824, USA;}
\affil{Department of Computational Mathematics, Science, and
  Engineering, Michigan State University, East Lansing, MI 48824, USA}
\affil{Facility for Rare Isotope Beams, Michigan State University, East Lansing, MI 48824, USA}

\author[0000-0002-8228-796X]{Evan P. O'Connor}
\affiliation{The Oskar Klein Centre, Department of Astronomy,\\ Stockholm University, AlbaNova, SE-106 91 Stockholm, Sweden}
	
\begin{abstract}

Gravitational waves (GWs) provide unobscured insight into the birthplace of neutron stars (NSs) and black holes in core-collapse supernovae (CCSNe). The nuclear equation of state (EOS) describing these dense environments is yet uncertain, and variations in its prescription affect the proto-neutron star (PNS) and the post-bounce dynamics in CCSNe simulations, subsequently impacting the GW emission. We perform axisymmetric simulations of CCSNe with Skyrme-type EOSs to study how the GW signal and PNS convection zone are impacted by two experimentally accessible EOS parameters, (1) the effective mass of nucleons, $m^\star$, which is crucial in setting the thermal dependence of the EOS, and (2) the isoscalar incompressibility modulus, $K_{\rm{sat}}$. While $K_{\rm{sat}}$ shows little impact, the peak frequency of the GWs has a strong effective mass dependence due to faster contraction of the PNS for higher values of $m^\star$ owing to a decreased thermal pressure. These more compact PNSs also exhibit more neutrino heating which drives earlier explosions and correlates with the GW amplitude via accretion plumes striking the PNS, exciting the oscillations. We investigate the spatial origin of the GWs and show the agreement between a frequency-radial distribution of the GW emission and a perturbation analysis. We do not rule out overshoot from below via PNS convection as another moderately strong excitation mechanism in our simulations.  We also study the combined effect of effective mass and rotation. In all our simulations we find evidence for a power gap near $\sim$1250 Hz, we investigate its origin and report its EOS dependence.

\end{abstract}

\section{Introduction}\label{sec:intro}

Core-collapse supernovae (CCSNe) are the extraordinary birth sites of neutron stars (NSs) and stellar-mass black holes (BHs) and are triggered from the gravitational collapse of a massive star. During a CCSN, a proto-neutron star (PNS) forms where the iron core of the massive star once was. In a PNS, which subsequently evolves into a NS or a BH, matter exists in a wide range of temperatures, $0 \lesssim T \lesssim \mathcal{O}(100\unit{MeV})$, can be compressed to densities up to and beyond $\rho \simeq 10^{15}\unit{g\,cm}^{-3}$, and may be proton rich, $y \simeq 0.6$, or very neutron rich, $y \simeq 0$, where $y$ is the ratio of protons to baryons. As some of these conditions are rarely found elsewhere in the Universe, PNSs are unique probes of matter in its most extreme states. However, PNSs are hidden from our direct sight by the outer layers of the massive star. 
Thus, only neutrinos and gravitational waves (GWs) produced in the aftermath of the core collapse of a massive star can provide unobscured insight into these extreme environments. 

As the recent detection of gravitational waves from merging binary systems of BHs \citep[e.g.][]{abbott:GW150914} and NSs \citep{abbott:GW170817} marks the onset of GW interferometer astronomy, hope arises that the next major breakthrough comes from a multimessenger event of a nearby CCSN \citep{szczepanczyk:21}. Since the estimated Galactic CCSN rate is 3$^{+7.3}_{-2.6}$ per century \citep{li:11, adams:13}, such hope is scientifically justified. The yield of knowledge from such an event is bounded by our understanding of CCSNe and the description of dense nuclear matter. The nuclear equation of state (EOS) is one of the fundamental ingredients for predicting the dynamics of CCSNe. The spectra of neutrinos, the PNS mass, its radius, cooling rate, and the GW signal are all dependent on the EOS. As details and parameters in the EOS of dense nuclear matter are yet poorly constrained \citep{oertel:17}, a wide range of EOSs are typically used in CCSNe simulations \citep{hempel:12}. As a result, there are several studies exploring the dependence of the GW emission on the nuclear EOS \citep[][]{kuroda:17,pan:18, morozova:18}, however, due to the large nuclear EOS model space available and the diversity of available EOSs in this model space, disentangling the impact of individual EOS parameters on the CCSN dynamics must be done carefully.

Despite differences in microphysics and model construction, some of these EOSs can be characterized in terms of semi-analytic, semi-experimental empirical parameters defined via a Taylor expansion about saturation density \citep{oertel:17,margueron:18}. This approach, referred to as meta-modeling, was used by \cite{schneider:19} to construct a set of finite-temperature EOSs based on the framework developed by \citet{lattimer:85, lattimer-swesty:91, schneider:17}. These EOSs describe nucleon interactions via a Skyrme force and the thermal contributions in the homogeneous phases, regions of parameter space where heavy nuclei do not appear, are fully determined by the effective mass of nucleons.

In a comprehensive set of spherically-symmetric CCSNe simulations (and also limited 3D simulations) of the 20-$M_\odot$ progenitor star of \citet{woosley-heger:07}, \cite{schneider:19} report an effective mass dependence of the post-bounce dynamics. As the effective mass serves as a proxy for the EOS temperature dependence, changing its value affects the thermal pressure throughout the PNS. In fact, increasing the value of the effective mass parameter leads to a decrease in thermal pressure in the core of the PNS. Less thermal support from the core renders the PNS more compact and its outer regions, where the neutrinosphere is located, hotter. A hotter and more compact neutrinosphere emits higher energy neutrinos with an increased luminosity and neutrino heating which drives stronger convection and turbulence in the gain region and facilitates successful SN explosions \citep{schneider:19,yasin:20}.  Higher turbulence and convection in the gain region has been suggested to impact PNS oscillations, and therefore potentially influence the amplitude of the GWs \citep{murphy:09,oconnor-couch:2018b}. Explosion dynamics aside, the compactness of the PNS is one of the main factors that determines the frequency of the GWs \citep{muller:13, pan:18, morozova:18}. Thus, a study of the thermal effects on the gravitational wave signal from the core-collapse of massive stars is opportune and is carried out here. We find a substantial impact of the effective mass on the peak GW frequency and its evolution through these aforementioned thermal effects.  

Recent efforts in understanding the nature of gravitational wave emission from CCSNe has also revealed, in some simulations, the presence of a ``power gap" in the frequency spectrum of the emitted gravitational wave signal.  As first noted by \cite{morozova:18}, the power gap is a narrow ($\Delta f \sim$ 50\,Hz), high frequency ($\sim 1200$\,Hz) persistent (100s of ms in duration) lack of GW power that is present from the onset of GW emission. \cite{morozova:18} suggest this gap has its origins in an interaction between trapped modes in the PNS core and higher frequency p-modes present in the outer parts of the PNS.  As such, it is likely linked to the eventual transition of the dominant GW emission from that of a core g-mode to a higher-frequency mode, either the f-mode \citep{sotani:2020,morozova:18} or another g-mode \cite{torres-forne:19b}. It remains to be seen whether this GW spectral feature is physical in nature, and what, if any, is its EOS dependence. We note, from the public availability of GW signals from CCSNe (with sufficient data to capture signals at $\sim$\,1300\,Hz), that the presence of this power gap is also seen in the 3D simulations of \cite{oconnor-couch:2018b} and \cite{radice:19}, but not in the 3D simulation of \cite{mezzacappa:20}. We report on the presence of the power gap in our simulations presented here and explore its nature.

Related to the question of the presence and origin of the GW spectrum power gap is the origin of the GWs themselves. There is tension in the literature regarding the location in PNS where the GWs are emitted.  By decomposing the gravitational wave signal into its radial dependence, \cite{mezzacappa:20} suggests the bulk of the emission in their 3D simulation arises from inside the PNS convection zone.  This is supported to some extent by the work of \cite{andresen:17}, who decompose the signal into the contribution from the PNS convection zone (and its overshoot layer), the convectively-stable PNS surface zone, and the gain region.  \cite{andresen:17} find in their 3D models that the emission was predominantly coming from the PNS convection and overshoot layer and little emission was coming from the stable PNS layer itself. However, they note that the majority of the emission comes from the small overshoot region directly at the base of this layer and at the top of PNS convection layer. Similar analysis in \cite{andresen:17} of axisymmetric (2D) models, shows the potential for (perhaps unphysically) large GW emission in the convectively-stable PNS layer, however, this is not always present in the 2D simulations they study and may be associated with the presence of a strong standing accretion shock instability (SASI). In this article, we perform an in-depth analysis of the spatial origin of the GW emission.  We find evidence for the peak GW emission occurring in the convective overshoot layer at the top of the PNS convection zone, supporting the findings of \cite{andresen:17}. However, we note that the global nature of the GW emission, whose radial profile agrees with the results of a perturbation analysis, suggests that a precise determination of the PNS excitation mechanism via the spatial location of the GW emission alone may not be possible.

In this paper, we perform 2D axisymmetric simulations of CCSNe to investigate the impact of the effective mass and the incompressibility modulus of the EOS on the post-bounce GW signal. We also study the combined effect of rotation and effective mass. With this suite of simulations we also investigate the origin of the power gap and the spatial origin of the GW emission.  The paper is organized in the following way. In \S~\ref{sec:methods} we outline the numerical setup, equation of states used, and how we extract the GW signal from our numerical simulations. In \S~\ref{sec:results} we present our results on the dependence of the GW emission on the effective mass (\S\ref{sec:results:meff}), rotation (\S\ref{sec:results:rot}), incompressibility (\S\ref{sec:results:ksat}), spatial origin (\S\ref{sec:results:source}.), and present our investigation of the power gap (\S\ref{sec:results:gap}). We summarize and conclude in \S\ref{sec:conclusions}.

\section{Methods}
\label{sec:methods}
\subsection{Numerical Setup}

Each stellar core collapse simulation we perform uses a progenitor with zero-age main sequence mass of 20\,M$_\odot$ from \citet{woosley-heger:07} and is started at the onset of collapse in our 2D domain. We utilize the FLASH (v.4) multiscale, multiphysics adaptive mesh refinement framework \citep{fryxell:00} modified for core-collapse in \cite{couch:13,couch-oconnor:14,oconnor-couch:18}. The grid setup is a 2D cylindrical geometry that spans $1 \times 10^{9}$\,cm in the radial direction and $\pm 1 \times 10^{9}$\,cm along the cylindrical axis. Our maximum block size is $2 \times 10^{8}$\,cm and we allow a total of 10 levels of mesh refinement, giving a minimum block size of $\sim 3.9$\,km and minimum grid zone spacing of $\sim 326$\,m. Beyond 80 km, the highest level of refinement is limited so as to only maintain an angular resolution of 0$^\circ$.82. We use standard Newtonian hydrodynamics and solve gravity by employing a modified general relativistic effective potential (case A) for the monopole contribution \citep{marek:06}, retaining spherical harmonic orders up to 16. We note that the use of an effective potential leads to, in general, a quantitative over-prediction of the GW frequency due to the inconsistency of the underlying Newtonian hydrodynamics \citep{muller:13,zha:20}.

We incorporate a multidimensional, multispecies and energy dependent neutrino transport following an M1 scheme outlined in \citet{oconnor-couch:18}. As such, we distinguish three species of neutrino i.e., electron neutrinos, $\nu_e$, electron anti-neutrinos $\bar\nu_e$, and the composite group of heavy lepton neutrinos/anti-neutrinos $\nu_x$. For each EOS employed, a set of neutrino opacities is generated using the neutrino transport library NuLib \citep{oconnor:15}. Neutrino opacities are computed for 12 logarithmically-spaced energy groups, the first group centered on 1\,MeV and the last one centered on $\sim$315\,MeV.  For the emission of electron type neutrinos and anti-neutrinos, we include electron and positron capture interactions on protons and neutrons, respectively, as well as electron capture on heavy nuclei following \cite{bruenn:85}.   For the emission of heavy lepton neutrino pairs we include emission from electron-positron annihilation to $\nu\bar{\nu}$ pairs following \cite{bruenn:85} and nucleon-nucleon bremsstrahlung following \cite{burrows:06,hannestad:98}.  We include these pair emissions in a parameterized way \citep{betranhandy:20}. We include isoenergetic scattering of neutrinos of all types on nucleons, nuclei, and alpha particles following \cite{bruenn:85}. For the charged-current interactions on nucleons as well as the neutral-current scattering interactions we include corrections for weak-magnetism following \cite{horowitz:02}.  We also include inelastic scattering of neutrinos on electrons following \cite{bruenn:85}.

For the simulations that include rotation, we map the progenitor model onto the cylindrical grid via an artificial rotation profile,
\begin{equation}\label{eq:rot}
\Omega(r) = \Omega_0 \Big[  1 + \left( \frac{r}{A}\right )^2 \Big]^{-1}\,,
\end{equation}
\noindent
where $r = \sqrt{z^2 + R^2}$  is the spherical radius for a given cylindrical radius $R$ and symmetry axis coordinate $z$, $\Omega_0$ is the central angular speed of the star, and $A$ the differential rotation parameter. Using this same scheme, \cite{pajkos:19} find that the parameter $A$ strongly correlates with the stellar core compactness \citep{oconnor-ott:11}, and calculate an optimal value of $A  = 1021$\,km for the compactness $\xi_{2.5}$=0.2785 of the 20\,M$_\odot$ progenitor used here.

\subsection{Equation of State}
We use five EOSs from \cite{schneider:19} who constructed 97 Skyrme-type EOSs using the open-source code SROEOS \citep{schneider:17}. \cite{schneider:19} take pair-wise combinations of 8 empirically accessible parameters and allow them to vary within their estimated uncertainties based on current nuclear physics constraints \citep{danielewicz:02, margueron:18}. The EOSs are based on the commonly used ``liquid-drop model" framework of \citet{lattimer:85, lattimer-swesty:91} where nucleon interactions are computed via a Skyrme force. SROEOS has some improvements with respect to the model of \citet{lattimer-swesty:91}; the most relevant to this paper being that SROEOS (1) allows simple variations of the effective mass of nucleons instead of fixing them at their vacuum values and (2) the ability to compute EOSs for any desired value of the isoscalar incompressibility modulus $K_{\mathrm{sat}}$ rather than limiting them to $180$, $220$, and $375\unit{MeV\,baryon}^{-1}$. 

In this paper, we focus on variations in two empirical parameters to determine their effect on GW signatures. First, we focus on the effective mass of symmetric nuclear matter at saturation density, $m^\star\,=\,m^\star_n(n\,=\,n_{\mathrm{sat}},\,y=1/2) \simeq m^\star_p(n_{\mathrm{sat}},1/2)$ \footnote{The small difference between $m^\star_n(\,n_{\mathrm{sat}},1/2)$ and $ m^\star_p(n_{\mathrm{sat}},1/2)$ is due to the small neutron proton mass difference.}, where $n$ is the baryon number density, $n_{\mathrm{sat}} = 0.155$\,fm$^{-3}$, $y$ the proton fraction, and the subscripts $n$ and $p$ denote neutrons and protons, respectively. Second, we adjust the isoscalar incompressibility modulus $K_{\mathrm{sat}}$. We motivate this choice of parameters by their seemingly largest effect on the PNS radius in \citet{schneider:19}, followed by the isovector incompressibility modulus $K_{\mathrm{sym}}$. We note that the effect on PNS structure of $m^\star$ is much more significant than that of $K_{\mathrm{sat}}$ and, thus, we expect a similar correlation in our results. 

In Skyrme-type models, the inverse of the nucleon effective masses have a simple linear relation with nucleon densities. They are computed from
\begin{equation}\label{eq:meff}
\frac{\hbar^2}{2m_t^{\star}} = \frac{\hbar^2}{2m_t} + \alpha_1 n_t+ \alpha_2 n_{-t}\, ,
\end{equation}
\noindent
where $\alpha_i$ are Skyrme parameters, $m_t$ is the vacuum mass, and $n_t$ are number densities of nucleon of type $t$. For $n_t$, if $t = n$ then $-t = p$ and vice versa. The parameters $\alpha_i$ are set to reproduce the desired effective mass behavior. We refer to \citet{schneider:19} for further details. Note however that the finite-temperature component of the EOS is sensitive to the effective masses of nucleons \citep{prakash:97,constantinou:14}, while the incompressibility modulus enters as a parameter in the Taylor expansion around saturation density which describes the zero-temperature component \citep{margueron:18,schneider:19}. Thus, while  $K_{\mathrm{sat}}$ ($m_t^\star$) has a strong (weak) effect on the \textit{cold} NS structure it has a weak (strong) effect on the structure of a hot PNS. 

As in \cite{schneider:19, schneider:20}, we begin with a simulation where the EOS parameters assume a set of baseline values (see Table~\ref{table:baseline}), in particular with $m^\star = 0.75$ (specified as a fraction of the neutron mass vacuum value) and $K_{\mathrm{sat}} = 230$\,MeV\,baryon$^{-1}$. Next, we keep all but one parameter at baseline, varying it with $2\sigma$ above and below baseline. The $2\sigma$ values for $m^{\star}$ ($K_{\mathrm{sat}}$) below baseline is 0.55 (200\,MeV\,baryon$^{-1}$) and above baseline is 0.95 (260\,MeV\,baryon$^{-1}$). For both $m^{\star} = 0.55$ and $m^{\star} = 0.95$, two additional simulations are performed including rotation with $\Omega_0 =$ 1\,rad\,s$^{-1}$ and $\Omega_0 =$ 2\,rad\,s$^{-1}$. See Table~\ref{table:simulations} for an overview of the nine simulations performed in this paper.

\begin{table}[tb]
     \centering
     \begin{tabular}{cccc}
     \hline \hline
       \multicolumn{1}{c}{\textbf{Quantity}} & \textbf{Baseline} & \textbf{Variation 2$\sigma$} & \textbf{Unit} \\ \hline
        $m^\star$                 & 0.75     & 0.20 &  $m_n$    \\
        $K_{\mathrm{sat}}$        & 230      & 30   &  MeV\,baryon$^{-1}$  \\\hline
        $n_{\mathrm{sat}}$        & 0.155    &      &  fm$^{-3}$  \\
        $\epsilon_{\mathrm{sat}}$ & -15.8    &      &  MeV\,baryon$^{-1}$  \\
        $\epsilon_{\mathrm{sym}}$ & 32       &      &  MeV\,baryon$^{-1}$  \\
        $L_{\mathrm{sym}}$        & 45       &      &  MeV\,baryon$^{-1}$   \\
        $K_{\mathrm{sym}}$        & -100     &      &  MeV\,baryon$^{-1}$   \\
        $\Delta m^\star$          & 0.10     &      &  $m_n$   \\
        $P^{(4)}_{\mathrm{SNM}}$  & 125      &      &  MeV\,fm$^{-3}$  \\
        $P^{(4)}_{\mathrm{PNM}}$  & 200      &      &  MeV\,fm$^{-3}$  \\ \hline
     \end{tabular}
     \caption{Nuclear matter properties of the baseline EOS, following \citet{schneider:20}. See \cite{schneider:19} for further details. 2$\sigma$ variations are shown for the two parameters we investigate in this work.}
     \label{table:baseline}
   \end{table}

\begin{table}[ht]
     \centering
     \begin{tabular}{ccccc}
     \hline \hline
       \multicolumn{1}{c}{\textbf{$m^\star$}} & \textbf{$K_{\mathrm{sat}}$} & \textbf{$\Omega_0$} & \textbf{Label}\\
        \multicolumn{1}{c}{[$m_n$]} & [MeV\,baryon$^{-1}$] &  [rad\,s$^{-1}$] & \\ \hline 
        0.75* & 230* & 0 & \texttt{m0.75/k230} \\ \hline
        0.55 & 230*  & 0 & \texttt{m0.55}    \\
        0.95 & 230*  & 0 & \texttt{m0.95}   \\ \hline
        0.55 & 230*  & 1 & \texttt{m0.55r1} \\
        0.95 & 230*  & 1 & \texttt{m0.95r1} \\ 
        0.55 & 230*  & 2 & \texttt{m0.55r2} \\
        0.95 & 230*  & 2 & \texttt{m0.95r2} \\ \hline
        0.75* & 200  & 0 & \texttt{k200}    \\
        0.75* & 260  & 0 & \texttt{k260}   \\ \hline
     \end{tabular}
     \caption{Overview of the simulations performed that vary in the effective mass parameter, rotation rate, and the isoscalar incompressibility modulus parameter. * values are the baseline from Table~\ref{table:baseline}.}
     \label{table:simulations}
   \end{table}
   
\subsection{Gravitational Wave Extraction}
To extract the GW signal measured by a distant observer, we adopt the standard formula utilizing the second time derivative of the trace-free quadrupole moment.  This quadrupole moment in the slow motion, weak-field formalism is given by\citep{blanchet:90, finn-evans:90},

 \begin{equation}\label{eq:quadrupole}
       I_{ij} = \int d^3x(x_i x_j - \frac{1}{3}x^2\delta_{ij} )\rho(t, \vec{x})\,,
 \end{equation}
where the indices range over $(x,y,z)$, $\delta_{ij}$ is the Kronecker delta and $\rho$ the source energy density. For axisymmetric sources, the only independent component of the quadrupole moment is $I_{zz}$. Unless otherwise stated (for example, in \S~\ref{sec:results:source}), we compute the first derivative in terms of the fluid velocity, $v_i$, via the analytic expression \citep{reisswig:11},
\begin{equation}
\frac{dI_{ij}}{dt} = \int d^3x (x_iv_j + v_ix_j - 2/3\delta_{ij}x_k v_k)\rho(t, \vec x)\,,\label{eq:dIijdt}
\end{equation}
\noindent
 and the second time derivative is computed numerically.   The axisymmetric GW strain for an observer located at the equator a distance $D$ away from the source is then given as \citep{blanchet:90, finn-evans:90},

 \begin{equation}\label{eq:radiation}
  h_{+}(t, \vec{x}) = \frac{2}{D}\frac{G}{c^4} 
  \frac{d^2}{dt^2} I_{zz}\,,       
 \end{equation}
\noindent
where $G$ is the gravitational constant, $c$ the speed of light.

The GW spectrograms are extracted by short-time Fourier transforming (STFT) the GW signal,
\begin{equation}
\tilde{S}(f, \tau) = \int_{-\infty}^{\infty} s(t) H(t-\tau) e^{-2\pi i f t} dt\,,
\end{equation}
\noindent
into its frequency domain $f$. $\tau$ is the time offset of the Hann window function $H(t - \tau)$ \citep[see e.g.][]{murphy:09}. The sampling frequency of our simulation output is $\sim 2\times10^6$\,Hz, which is reduced in the post-processing to $\sim 20,000$\,Hz. Unless otherwise stated, we use a Hann window width of 40\,ms and compute STFTs with a $\Delta \tau$ of 3.6\,ms.

To investigate the hypothetical detectability of our simulations, we calculate the dimensionless characteristic strain \citep{murphy:09},

\begin{equation}\label{eq:hchar}
h_{\mathrm{char}} = \sqrt{ \frac{2}{\pi^2}  \frac{G}{c^3} \frac{1}{D^2} \frac{dE_{\mathrm{GW}}}{df} }\,,
\end{equation}
\noindent
with the spectral energy density,

\begin{equation}
\frac{dE_{\mathrm{GW}}}{df} = \frac{3}{5}\frac{G}{c^5} (2 \pi f)^2 \tilde{| A |}^2\,, 
\end{equation}

\vspace*{0.5\baselineskip}
\noindent 
where $\tilde{A}$ denotes the Fourier transform of $d^2I_{zz}/dt^2$.

\section{Results}
\label{sec:results}

In this section, we describe the main results of our study.
We first present results for three axisymmetric CCSNe simulations that show how the effective mass affects the PNS evolution and GW signal for the non-rotating $20\,M_\odot$ progenitor of \cite{woosley-heger:07}. 
Then we explore the effect of two rotation rates, $\Omega_0=1$ and $2\unit{rad\,s}^{-1}$, in the GW signal considering the two extreme cases studied for the effective mass. Finally, as a matter of interest and to reaffirm the findings of \cite{schneider:19}, we explore the impact of the isoscalar incompressibility, $K_{\rm sat}$, within its current experimental constraint. Recall that we label the simulations following the convention in Table~\ref{table:simulations}. Following this, we explore the spatial origin of the GWs in our models and the power gap. All of the GW wave strains reported in this work are available at \url{https://doi.org/10.5281/zenodo.4973546}.

\subsection{Effective Mass}
\label{sec:results:meff}
\begin{figure}
\begin{center}
\includegraphics[width=\linewidth]{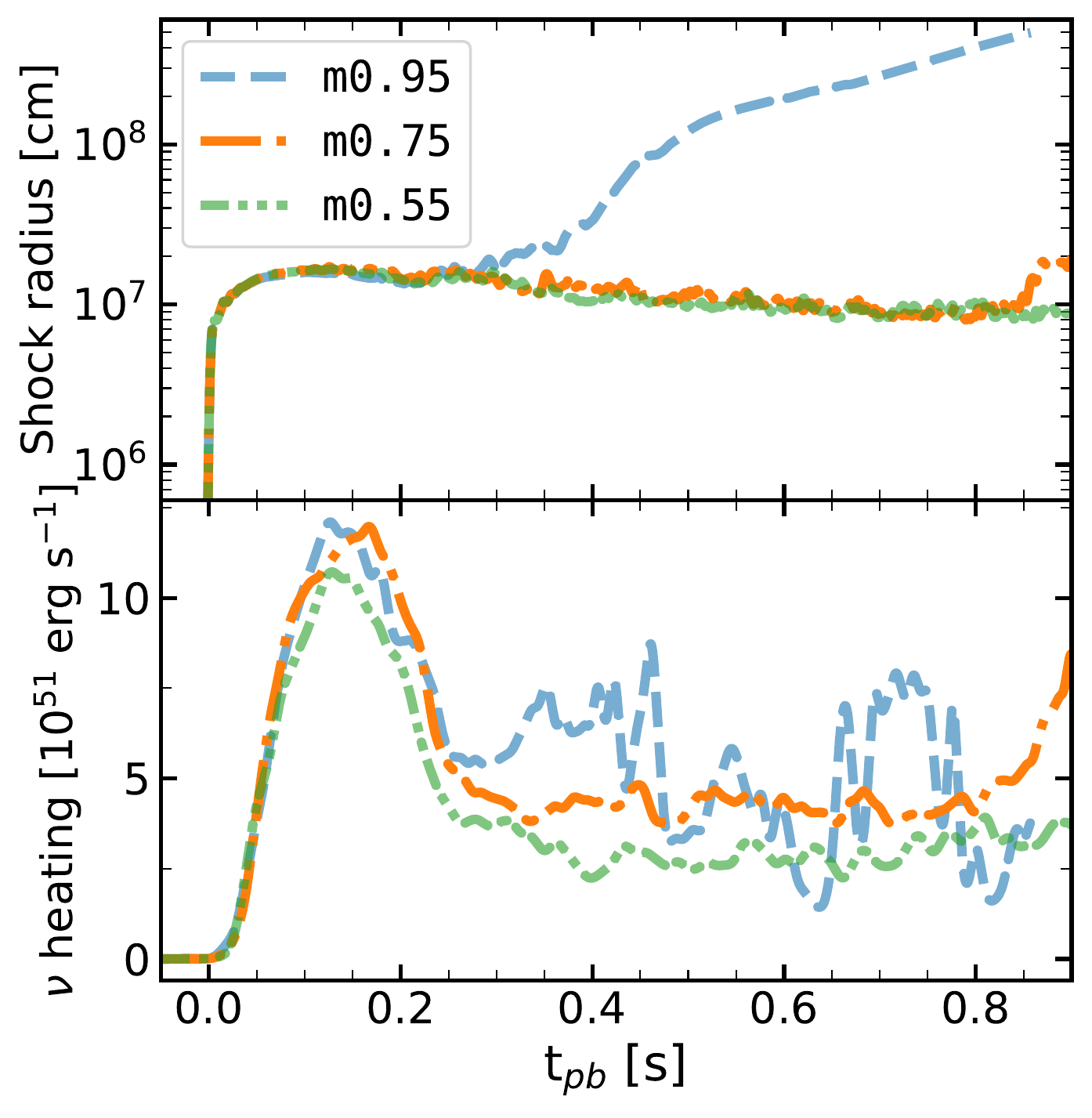}
\caption{Evolution of the mean shock radius (\textit{top panel}) and the neutrino heating (\textit{bottom panel}). Shock revival occurs for \texttt{m0.95} at $\sim 300$\,ms post-bounce, and for \texttt{m0.75} at $\sim 900$\,ms post-bounce, whereas the simulation with the lowest value of the effective mass parameter does not explode in the simulated time. This is generally consistent with the increased electron-type neutrino luminosity and mean energies and therefore neutrino heating for the simulations with the higher effective mass, making conditions more favorable for shock revival.}
\label{fig:meff_descr}
\end{center}
\end{figure}

\begin{figure*}[htb]
\begin{center}
\includegraphics[width=0.8
\linewidth]{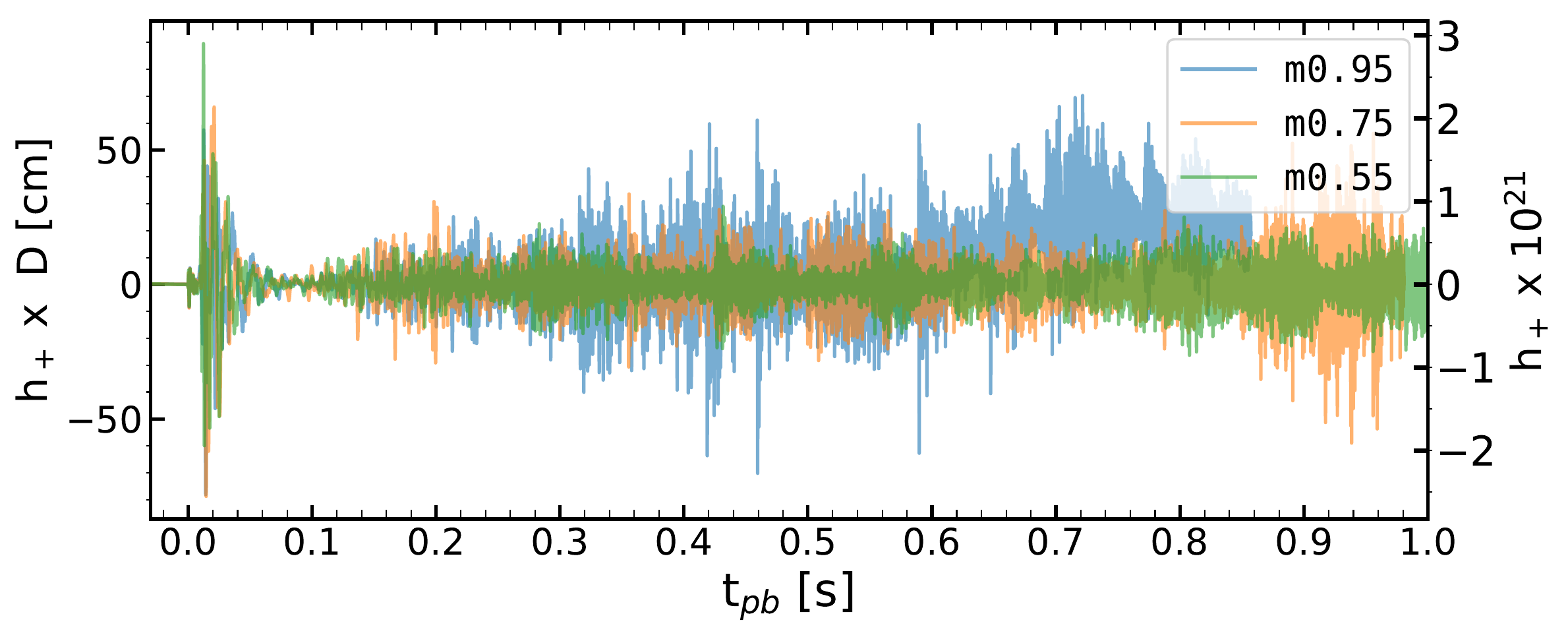}
\caption{Gravitational wave signals for models varying in the effective mass parameter as measured from an arbitrary distance (left axis) and from 10\,kpc (right axis), assuming that the observers are standing in the equatorial plane of the source. Generally, the GW amplitude increases with a larger value of the effective mass parameter. Around 0.6\,s, the signal for the \texttt{m0.95} model undergoes a shift in the positive direction due to a prolate shape of the shock during explosion.}
\label{fig:hxd_meff}
\end{center}
\end{figure*}

\begin{figure}[ht]
\begin{center}
\includegraphics[width=\linewidth]{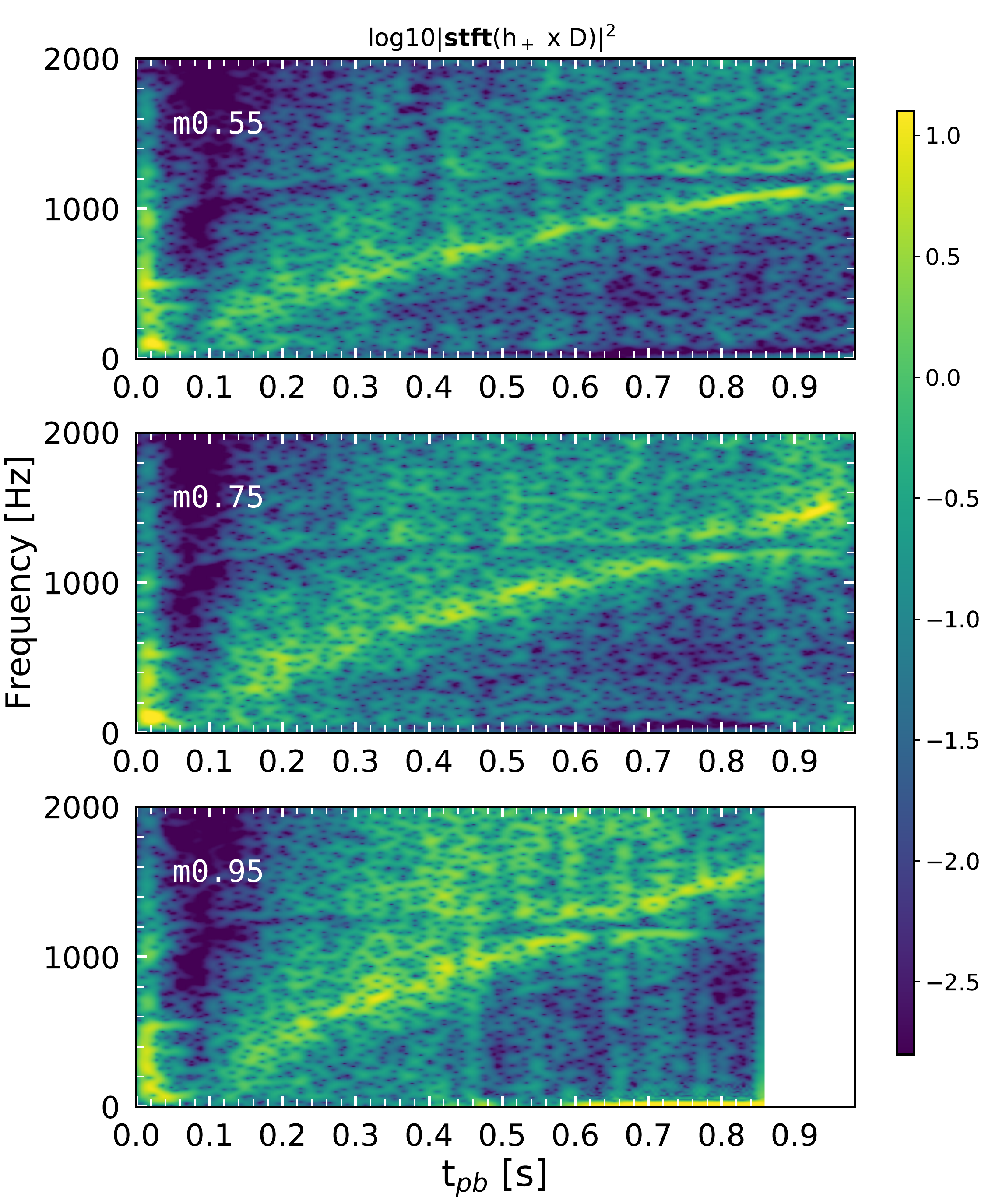}
\caption{Logarithmic spectrograms of the gravitational wave signal for the three simulations varying in the effective mass. For a direct comparison, they are normalized across their global maximum above 100\,ms. There are three distinct features in the spectrograms: (1) strong emission over a large range of frequencies $\lesssim 70$\,ms typically attributed to prompt convection, (2) a narrow band of high power that increases in frequency with time, the dominant PNS oscillation frequency, and (3) a region located at approximately constant frequency ($\sim 1.2$\,kHz) that appears to lack emission, the ``power gap" which we explore in \S\ref{sec:results:gap}.}
\label{fig:spec_meff}
\end{center}
\end{figure}

\begin{figure}[ht]
\begin{center}
\includegraphics[width=\linewidth]{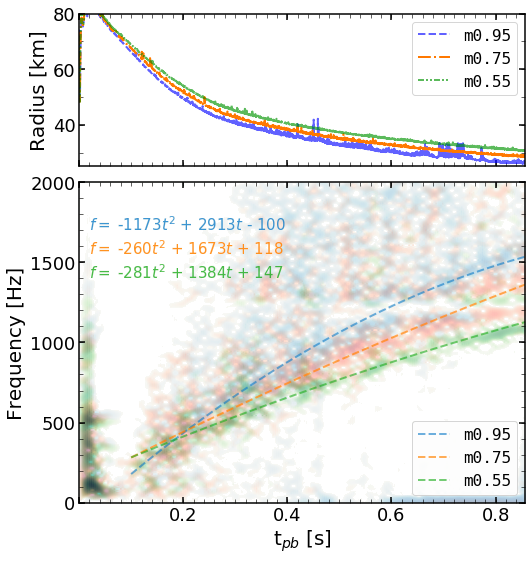}
\caption{\textit{Top panel}: PNS radius defined as the maximum radius where the density is at least $10^{11}$\,g\,cm$^{-3}$. \textit{Bottom panel}: Overlaid spectrograms of the three models varying in effective mass and overplotted with second order polynomial fits to the frequency points of most power. The window over which the fit is made is from 0.1\,s to the end of the simulation time. The polynomial coefficients are shown in the top left corner.}
\label{fig:spec_overlay_meff}
\end{center}
\end{figure}

When investigating the effective mass dependence of GWs for non-rotating progenitors, the three simulations we compare are carried out for a minimum of 860\,ms post-bounce. Each stellar core collapses until the EOS stiffens and bounce occurs (at $318.5 \pm 0.1 $\,ms after the onset of the simulation), followed by the formation and outward propagation of the shock wave that initially stalls at $\sim 170$ km at $\sim$100\,ms after bounce. The top panel of Figure~\ref{fig:meff_descr} shows the evolution of the mean shock radius. Shock revival occurs for \texttt{m0.95} at $\sim 300$\,ms post-bounce, and for \texttt{m0.75} at $\sim 900$\,ms post-bounce, whereas the simulation with the lowest value of the effective mass parameter, \texttt{m0.55}, does not explode in the simulated time. This is consistent with the generally increasing neutrino heating expected for higher effective masses (bottom panel of Figure~\ref{fig:meff_descr}), making conditions more favorable for shock runaway \citep{janka:17, yasin:20, schneider:19}. The difference in neutrino heating between the simulations is a reflection of the neutrino luminosity  and neutrino mean energy, both of which increase with the effective mass. We remark that this trend is seen for all neutrino/anti-neutrino species with eventual deviations related to the times of explosion. 

Neutrino heating does not only provide the shock with thermal support and aid in shock revival, it also drives turbulence and convection in the gain region \citep{couch-ott:15} that imprints on the gravitational waves. Figure~\ref{fig:hxd_meff} shows the gravitational wave signals. While we note a strong signal at epochs $\lesssim 70$\,ms that is typically attributed to prompt convection \citep{murphy:09}, our analysis focuses on epochs after 0.1\,s when the quiescent phase has ended and the excitation of PNS oscillations begins. During this phase, such PNS oscillations strongly dominate the GW signal. The amplitude of the gravitational waves increases with higher effective mass.
We attribute this, at least in part, to the increasingly (neutrino driven) ``violent" gain region in simulations with higher effective mass, correlating with the amplitude of the PNS oscillations via accretion plumes striking the convectively stable layer below \citep{murphy:09, yakunin:15}. \citet{radice:19} shows, albeit in 3D, that the energy radiated in GWs is proportional to the amount of turbulent energy accreted by the PNS. Here, we note (but do not show) that the total energy radiated increases with the effective mass throughout all evolutionary stages above 0.1\,s. It is worth noting that the development of an explosion, as long as there is sustained accretion, can lead to GWs with an amplitude at a level higher than it would have been without an explosion \citep{radice:19}, therefore, the high GW amplitude seen in \texttt{m0.95}, and to some extent in \texttt{m0.75}, after the onset of explosion can be in part attributed with this as well.
We return with an in-depth analysis and discussion regarding the source of the GWs in \S\,\ref{sec:results:source}.

As seen in previous studies \citep[e.g.][]{murphy:09, yakunin:15}, we observe a prolate shape of the shock for the \texttt{m0.95} model which is reflected in the GW signal by a shift in the positive direction at $\sim 0.6$\,s.\\

In Figure~\ref{fig:spec_meff}, we show the spectrograms where the typical, dominant PNS oscillation GW signal starts at $\sim 0.1$\,s and increases in frequency as the PNS cools and contracts. In all three spectrograms, we note a particular region that appears to lack emission, located at roughly constant frequency ($\sim 1.2$\,kHz) throughout the simulation time. \citet{morozova:18} were the first to address this ``power gap", which they could not attribute to a numerical artifact, and, thus, do not rule out a physical origin. As the dominant oscillation mode evolves towards the gap region from below, an interaction appears to take place with another mode above the power gap, consistent with the avoided-crossing description in \cite{morozova:18,sotani:2020}, resulting in the lower frequency  mode flattening out in frequency and dropping in amplitude, while the other higher frequency mode rises in frequency and amplitude. The crossing between modes occur at roughly $\sim 650$\unit{ms}, $\sim 850\unit{ms}$, and $\sim 1000\unit{ms}$ for the \texttt{m0.95}, \texttt{m0.75}, and \texttt{m0.55} simulations, respectively. We report our further exploration of the power gap in \S\ref{sec:results:gap}.

\begin{figure}[ht]
\begin{center}
\includegraphics[width=\linewidth]{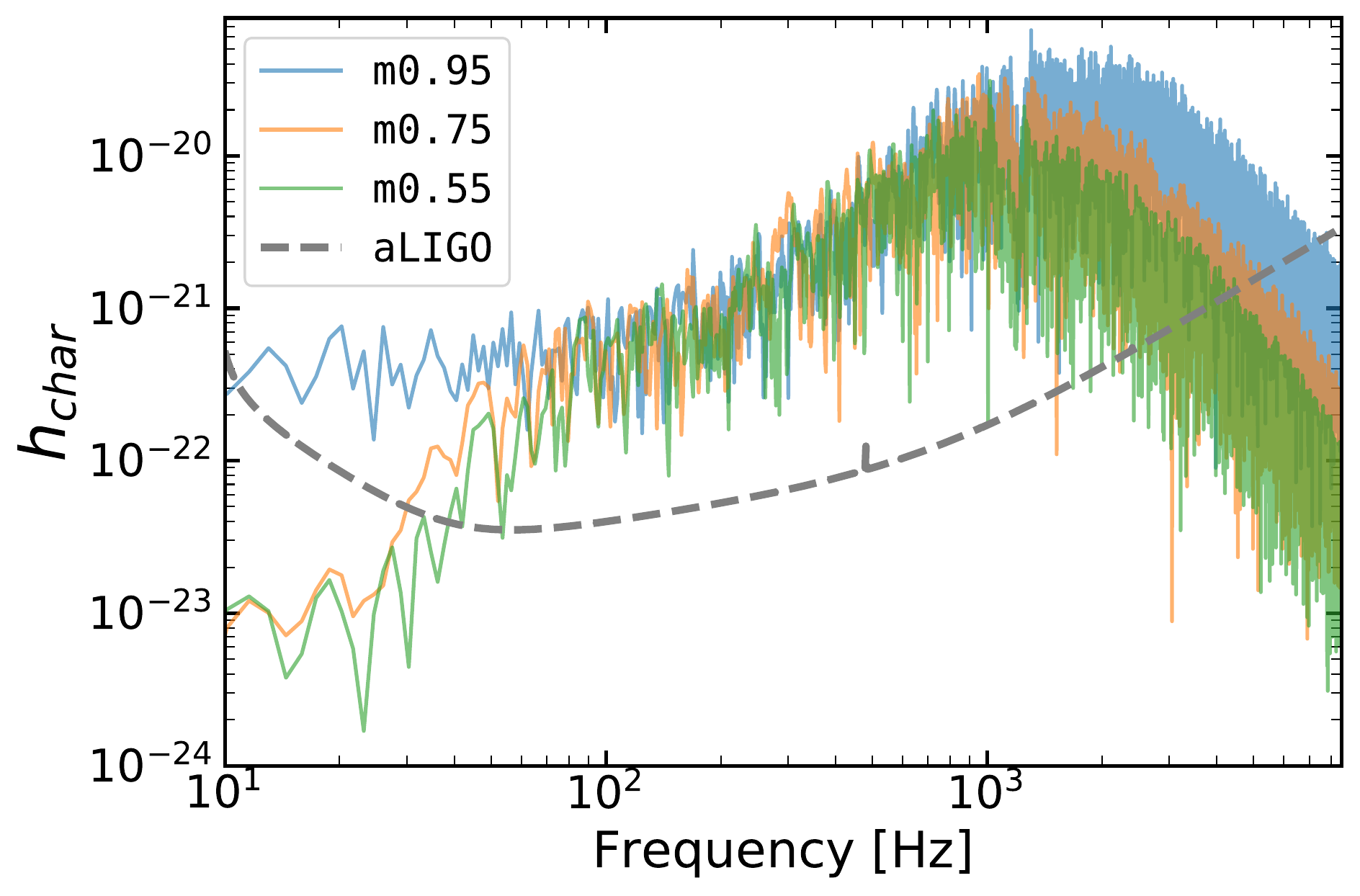}
\caption{Characteristic strain as a function of frequency, assuming a distance to the source of 10 kpc. This is calculated between $100-860$\,ms post-bounce. The peak frequency and its corresponding strain increases as the value of the effective mass increases. The grey dashed line represents the sensitivity bounds of Advanced LIGO \citep{sensitivity:aLigo}, indicating potential of detecting gravitational wave signals from a Galactic CCSNe of similar physical properties as these models.}
\label{fig:meff_hchar}
\end{center}
\end{figure}

We show in Figure~\ref{fig:spec_overlay_meff} the radius of the PNSs (top panel) and the three spectrograms overlaid on one another (bottom panel). We define the PNS radius as the maximum distance to where the density is $10^{11}$\,g\,cm$^{-3}$ which leads to sporadic jumps in the evolution profile of the early-exploding run, \texttt{m0.95}. The dashed lines in the spectrogram plot are quadratic fits of the form $At^2 + Bt + C$ to the collection of frequency points of most power (one extracted per STFT time-step of 3.6\,ms). The PNS radius and the frequency of the dominant oscillation mode are clearly effective mass dependent. It has been realized through asteroseismology studies \citep[e.g.][]{morozova:18, torres-forne:19b} and works using semianalytical fits to the dominant frequency \citep[e.g.][]{muller:13, pan:18, pajkos:19}, that the mass and radius are the primary determinants of the dominant PNS oscillation frequency. The masses of the accreting PNSs remain equal between our models until $\sim 0.5$\,s after bounce and differ by at most 4\% by the end of run \texttt{m0.95} (at $t_{\rm{pb}}=860$\,ms) due to explosion, at which point the radii differ by $\sim15$ \% between our two extreme cases. Hence, the radius holds the dominant role in setting the frequency and accounts for the $\sim36$\,\% difference in frequency at 860\,ms between run \texttt{m0.55} and \texttt{m0.95}. As expected due to the role of the effective mass on the thermal properties outlined in \citet{schneider:19} and mentioned above, we find that a higher value of the effective mass parameter $m^\star$ yields a more compact PNS, producing a higher oscillation frequency.

Figure~\ref{fig:meff_hchar} shows the characteristic strain, $h_{\mathrm{char}}$ (Eq.~\ref{eq:hchar}) where we assume a distance of 10\,kpc to the source. The frequency corresponding to the largest $h_{\mathrm{char}}$, the so called peak frequency, is shifted to higher frequencies and higher power when increasing the effective mass, \textit{i.e.}, when increasing the PNS compactness. The peak frequencies are located around 1-2\,kHz, indicating that these amplitudes are due to the PNS oscillations. The power gap near $1.2\unit{kHz}$ is also clearly seen in this plot. Furthermore, because the peak frequency rises slower in the \texttt{m0.55} run and the window over which $h_{\mathrm{char}}$ is calculated is limited to the maximum common run time, $t_{\rm{pb}} =860$\,ms, the characteristic strain for this run above the power gap is lower than below the gap. This is not the case for the \texttt{m0.75} and \texttt{m0.95} runs, where the peak frequency crosses the power gap $\simeq200\unit{ms}$ earlier. 

Another feature we observe is a small shoulder of excess emission around $\sim 100\unit{Hz}$ which may be attributed to the  standing accretion shock instability (SASI) or more generally, neutrino-driven convection and turbulent flows in the gain region, where the GW radiation is typically emitted at such low frequencies \citep[e.g.][]{cerda-duran:13, pan:18, pajkos:19}. For the \texttt{m0.95} run there is also an excess emission at frequencies $\lesssim100\unit{Hz}$ due to the asymmetric explosion. 

The grey dashed line in Figure~\ref{fig:meff_hchar} represents a sensitivity curve of Advanced LIGO \citep{sensitivity:aLigo}. The frequency and amplitude ranges are well within the bounds of the sensitivity curve. Caution should be taken when concluding their detectability since 2D simulations typically overpredict the signal strength compared to 3D simulations. Several 3D studies report on amplitudes decreasing by a factor between $10-20$ in their 2D/3D comparisons \citep{andresen:17, oconnor-couch:18, mezzacappa:20}, which would make a detection at 10\,kpc more marginal, for example see \cite{szczepanczyk:21}.

\subsection{EOS and Rotation}
\label{sec:results:rot}
\begin{figure}[t]
\begin{center}
\includegraphics[width=\linewidth]{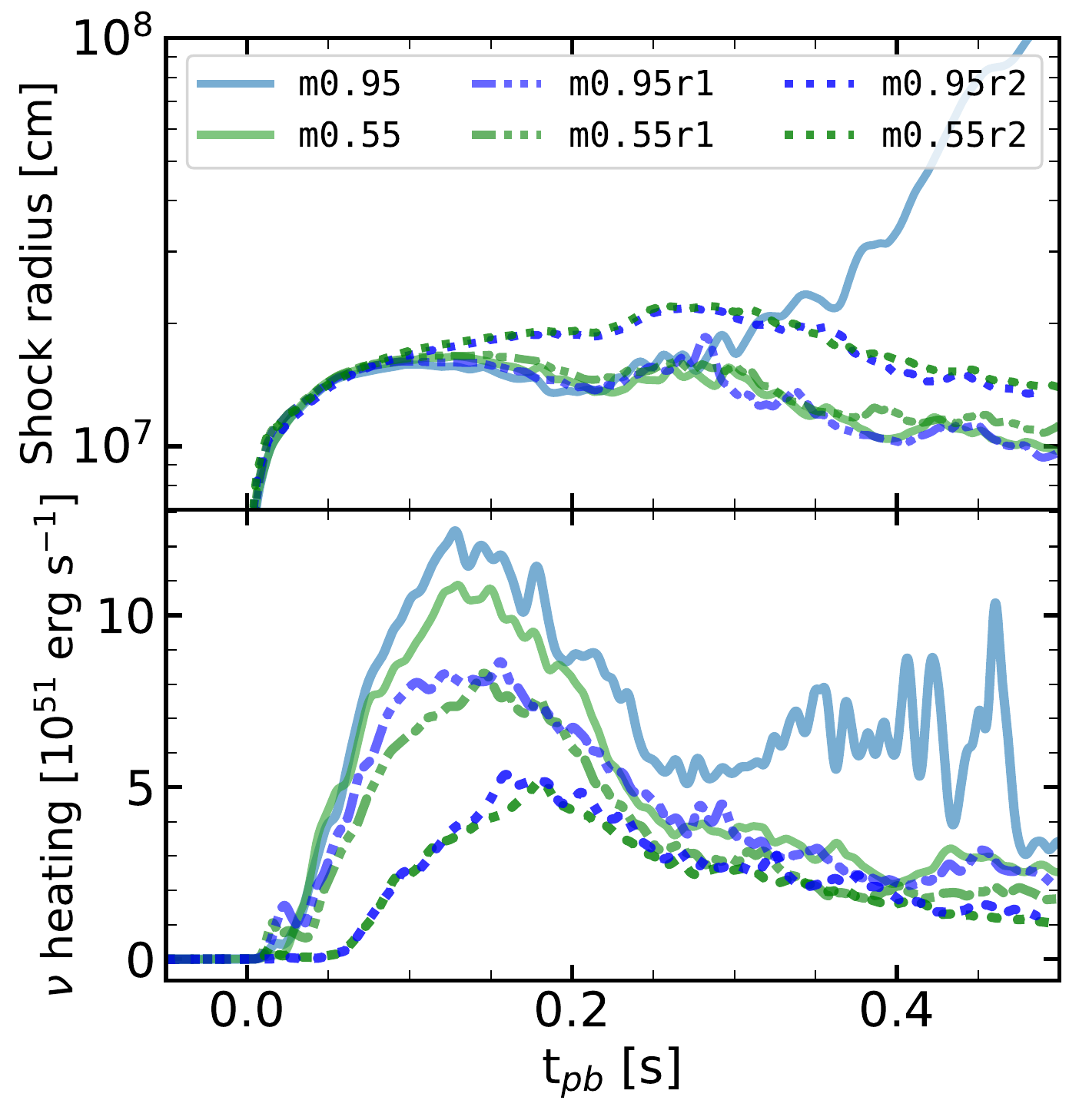}
\caption{\textit{Top panel}: Evolution of the mean shock radius comparing non-rotating models (solid lines), $\Omega_0=1\unit{rad\,s}^{-1}$ rotation (dashed-dotted lines), and $\Omega_0=2\unit{rad\,s}^{-1}$ rotation (dotted lines) for the $m^\star=0.95$ (blue) and the $m^\star=0.55$ (green) EOSs.  When rotation is included, no shock revivals occur during the $\sim0.5$\,s simulation time. \textit{Bottom panel}: Neutrino heating. Due to centrifugal support, rotating progenitors produce less compact PNSs, releasing less gravitational binding energy in the form of neutrinos and impacting the neutrino heating.}
\label{fig:rot_shock_heating}
\end{center}
\end{figure}

\begin{figure*}[ht]
\begin{center}
\includegraphics[width=0.9\textwidth]{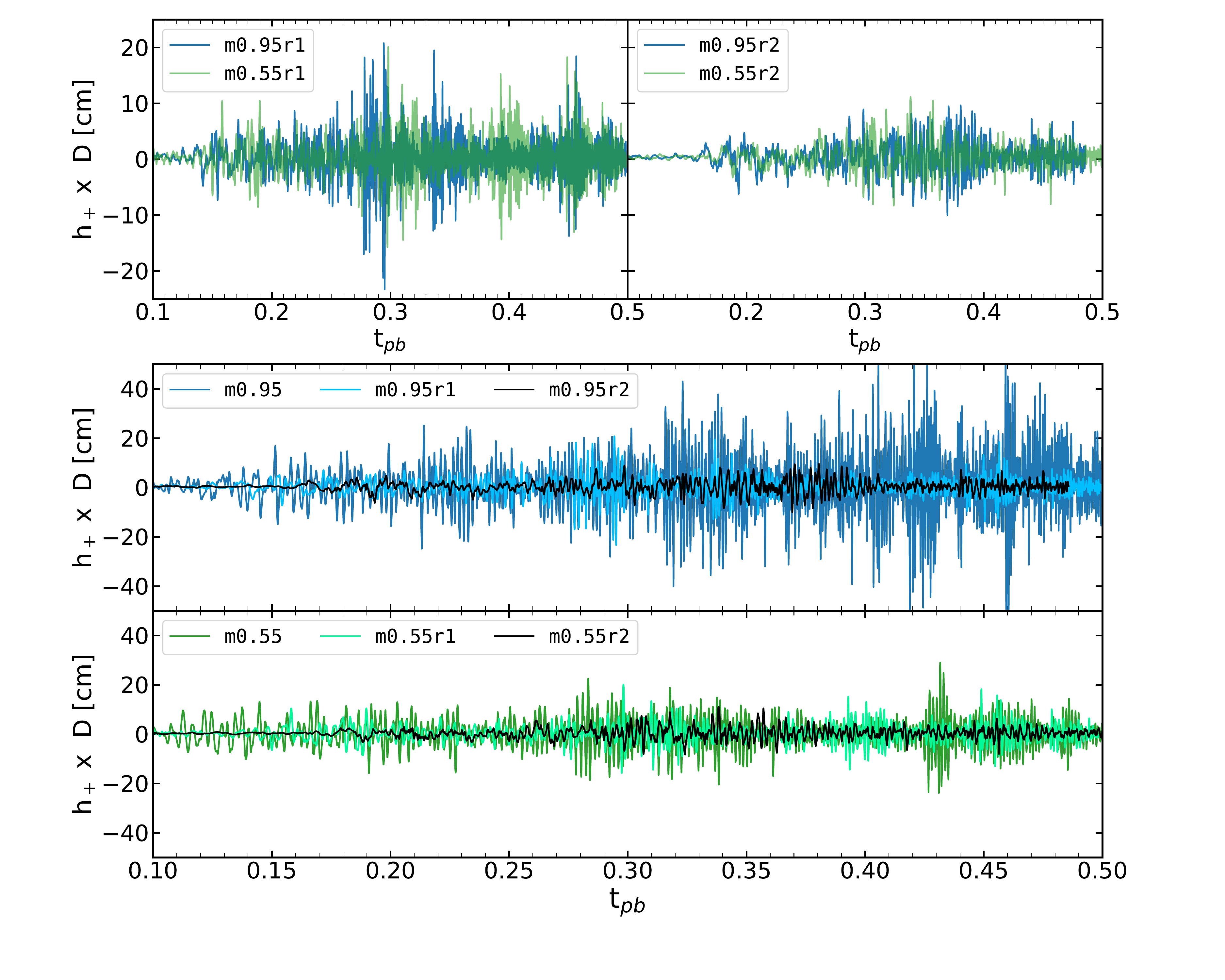}
\caption{GW signal for the models varying in both effective mass and rotation rate. \textit{Top left panel}: models with $\Omega_0 = 1$\,rad\,s$^{-1}$. \textit{Top right panel}: models with $\Omega_0 = 2$\,rad\,s$^{-1}$. \textit{Middle panel}: the $m^\star = 0.95$ models. \textit{Bottom panel}: the $m^\star = 0.55$ models. The two bottom panels emphasize how the signal becomes increasingly muted with higher rotation rate. The two panels on top show that the signal is muted to similar amplitude for equal rotation rates. Thus, the effective mass dependence of the GW amplitude is washed out by the somewhat centrifugally stabilized gain region.}
\label{fig:hxd_rot}
\end{center}
\end{figure*}

For our models with the highest and lowest values of the effective mass parameter, we run additional simulations using rotation profiles with central angular speeds of 1\,rad\,s$^{-1}$ and 2\,rad\,s$^{-1}$ at the onset of collapse (see Eq.~\ref{eq:rot}). Figure~\ref{fig:rot_shock_heating} shows the mean shock radius (top panel) and the neutrino heating (bottom panel). In contrast to the non-rotating \texttt{m0.95} simulation, no shock revivals occur for either \texttt{m0.95r1} nor \texttt{m0.95r2} during their $\sim 0.5$\,s simulation time. 

Due to centrifugal support, matter does not settle as deeply in the gravitational potential, releasing less gravitational binding energy during core collapse. This results in lower neutrino luminosity and slower contraction of the PNS \citep{westernacher:19,summa:18}. When we increase the rotation rate, we observe a significant non-linear decrease of the neutrino/anti-neutrino luminosities (as well as mean energies) which is reflected in the neutrino heating (Figure~\ref{fig:rot_shock_heating}). Other processes may factor into the observed heating also: for the most rapidly rotating models, the centrifugally supported shocks are located at a large radius, increasing the region where neutrinos may deposit energy. On the other hand, when including rotation, convection is inhibited due to a positive angular momentum gradient which is known to result in less prominent neutrino heating. \citep{murphy:13}.\\

Figure~\ref{fig:hxd_rot} shows the GW signal for the models varying in both the effective mass and rotation rate. Expectedly \citep[e.g.][]{pajkos:19}, the two bottom panels illustrate how the signal becomes increasingly muted with a higher rotation rate. As demonstrated by the two neighboring panels on top, and contrary to the non-rotating simulations, we do not find any effective mass dependence of the GW amplitude for our simulations that include rotation. This is of special note for the 1\,rad\,s$^{-1}$ models as they still exhibit an effective mass dependence in neutrino heating.  The centrifugal support stabilizes the gain region to the extent that the different amounts of neutrino heating has little to no impact on the GW amplitude. Furthermore, it takes longer for the convection instabilities in the 2\,rad\,s$^{-1}$ models to become fully non-linear in the gain region and thus these simulations exhibit an extended quiescent phase until $\sim 150-170$\,ms. Since these simulations are axisymmetric, the suppression of GW emission we see here due to rotation could be overestimated as compared to 3D. Furthermore, for sufficiently high rotation rates one expects non-axisymmetric instabilities give rise to potentially strong GWs \citep{scheidegger:10, andresen:18, takiwaki:18}.

\begin{figure}
\begin{center}
\includegraphics[width=\linewidth]{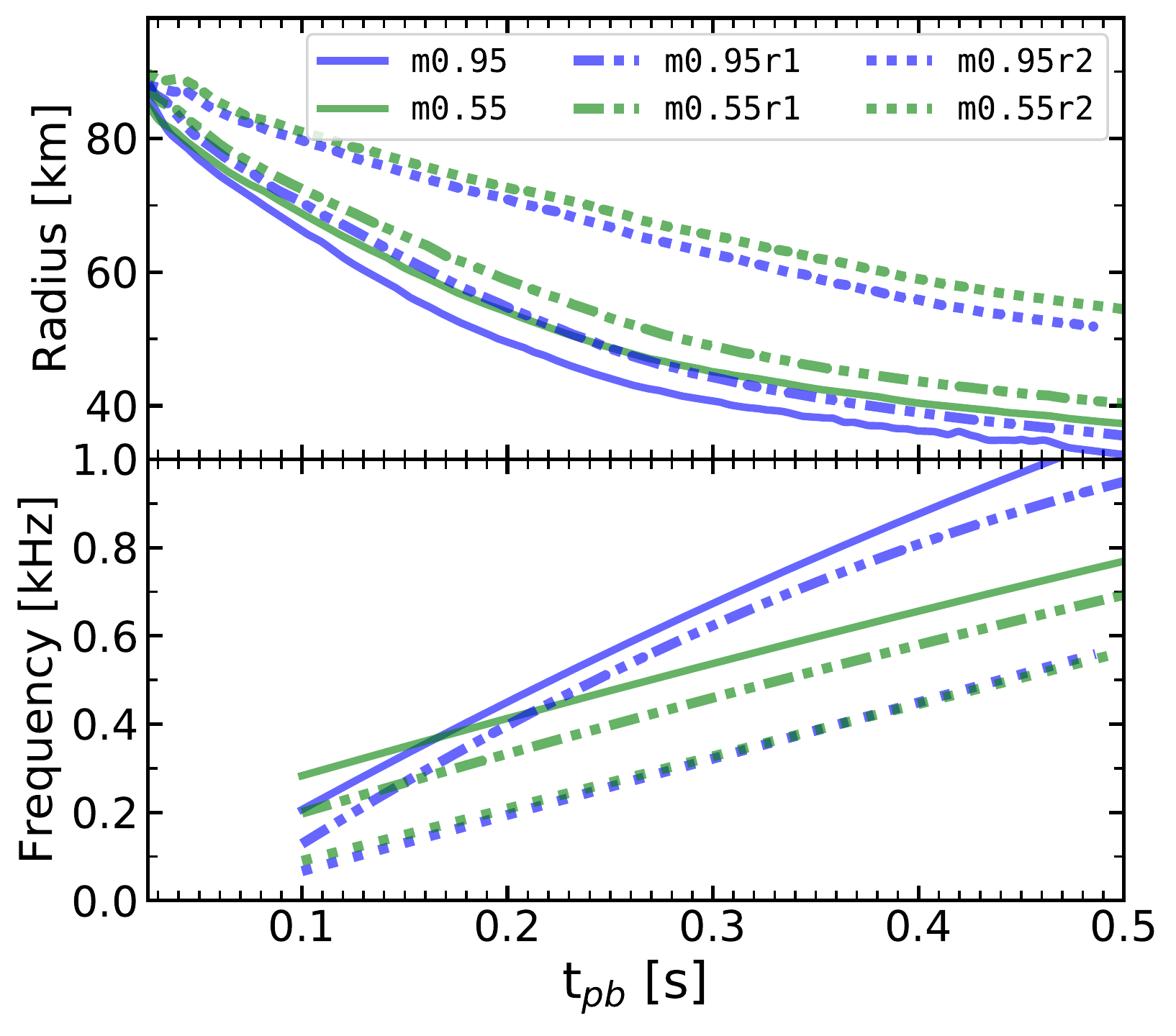}
\caption{The PNS radius (\textit{top panel}) for the models varying in both effective mass and rotation rate, as well as the polynomial fits of the form $At^2 + Bt + C$ to the frequency points of most power (\textit{bottom panel}). These frequencies are extracted from the spectrograms (one per time-step of 3.6\,ms) and we only use frequencies above 120\,Hz. For the 2\,rad\,s$^{-1}$ models, we use a linear fit due to a less well-defined dominant mode in the spectrogram (see text). The effective mass dependence of the dominant PNS oscillation frequency is still seen for the 1\,rad\,s$^{-1}$ models, but not the 2\,rad\,s$^{-1}$ models.} 
\label{fig:maxf_rot}
\end{center}
\end{figure}

Figure~\ref{fig:maxf_rot} shows the PNS radii (top panel) as well as the polynomial fits to the frequency points of most power (bottom panel). Correlated with the radius (that increases non-linearly with higher rotation rate), the frequency of the dominant PNS oscillation mode decreases with increasing rotation. The effective mass dependence of this mode is still clear for the 1\,rad\,s$^{-1}$ simulations, with rotation at this level only reducing the peak frequency by $\sim$10\% from the respective non-rotating counterparts. For the 2\,rad\,s$^{-1}$ models, due to the relatively little PNS excitation (and late onset thereof, see Figure~\ref{fig:hxd_rot}), the frequency distribution of power is much more broad and at a lower level.  Nevertheless, using a linear fit to the frequency points of most power for these rapidly rotating models (see the bottom panel of Figure~\ref{fig:maxf_rot}), we see a further reduction in the peak frequency and remark that the rapid rotation washes out any discernible frequency dependence that the slightly different PNS radii of the fast rotating simulations would cause.

\subsection{Incompressibility}
\label{sec:results:ksat}
The results above highlight how the compactness of the PNS at the time of formation as well as its cooling rate are the underlying factors of the GW signatures (under otherwise similar conditions). Indirectly so for the GW amplitude via the neutrino emission, and a more so direct correspondence between the PNS radius and the peak GW emission frequency.

\begin{figure}[ht]
\begin{center}
\includegraphics[width=\linewidth]{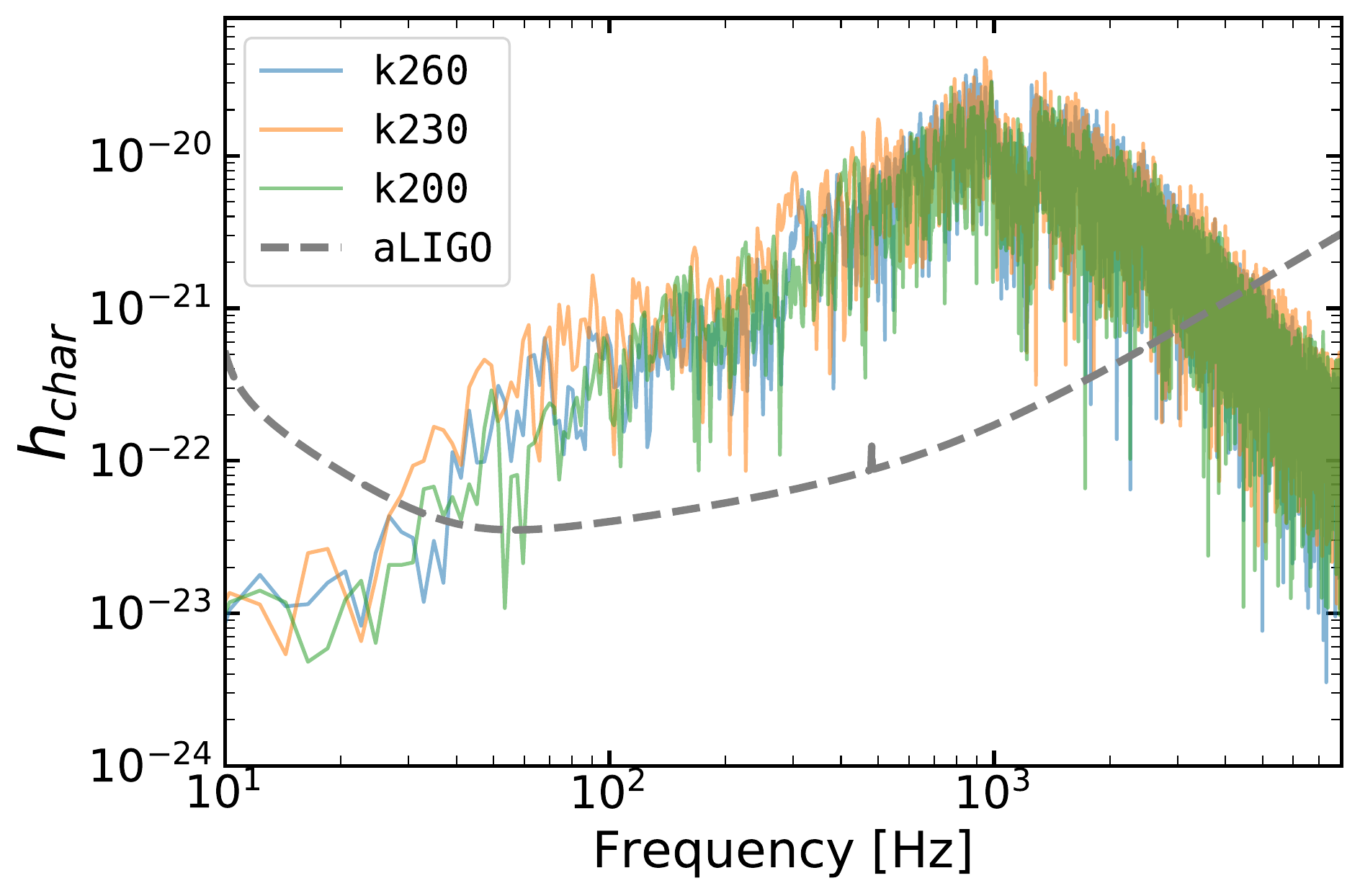}
\caption{Characteristic strain as a function of frequency for the models varying in the incompressibility modulus, $K_{\mathrm{sat}}$, assuming a 10 kpc distance to the source. This is calculated between $100-580$\,ms post-bounce. The grey dashed line represents the sensitivity bounds of Advanced LIGO \citep{sensitivity:aLigo}. There is little to no dependence of the GW signal on the incompressibility modulus.}
\label{fig:Ksat_hchar}
\end{center}
\end{figure}

For the simulations where we investigate the effect of the incompressibility modulus, the PNS masses and radii are nearly identical across the models during their $\sim 0.6$\,s simulation time. This agrees the spherically symmetric results of \cite{schneider:19}. Expectedly, we see no systematic differences in the GW signatures between the three models, and any minor differences are likely stochastic. Figure~\ref{fig:Ksat_hchar} shows the characteristic strain as a function of frequency, which reflects the similarity in both amplitude and frequency between the three simulations. 
We note that the decrease in GW emission in the $1.0$ to $1.2\unit{kHz}$ range is even more evident in Figure~\ref{fig:Ksat_hchar} now that the characteristic strains of the simulations $h_{\rm char}$ overlap. We discuss the incompressibility dependence of the gap further in \S\ref{sec:results:gap}.

\subsection{Source of the GWs}
\label{sec:results:source}
In a typical PNS there are two convectively stable regions, one in the inner core and one that composes the outer shell of the PNS that faces the turbulent post-shock region. According to canonical wave theory \citet{handler:13} (also see \cite{gossan:20} for a direct application in CCSNe), apart from oscillatory p-modes (where pressure acts as restoring force) at high frequency, these stable regions may also house g-modes at lower frequencies (where buoyancy acts as restoring force). These excitations may emit GWs at select eigenmode frequencies when excited. Often, a buoyantly unstable region separates the core and the outer shell, the PNS convection zone. In 2D, many groups have attributed their GW signal to g-modes from the outer shell \citep{cerda-duran:13, muller:13, pan:18}. A viable perturbing mechanism is accretion plumes impinging onto the PNS from the violent gain region above \citep{murphy:09}, also recognized in 3D \citep{oconnor-couch:18, radice:19, powell-muller:19}.

Adding to the possibilities of strong GW sources, \citet{mezzacappa:20} conclude that after a few 100s of ms, PNS convection takes over as the dominant source of GW emission in their 3D study. \citet{andresen:17} report that the dominant source of emission in their 3D runs originates from the bottom end of the convectively stable surface layer, driven by convective overshoot from the convection zone below rather than by accretion from above, and further show the same scenario for a 2D counterpart to a 3D run. It should also be mentioned that PNS convection has previously been recognized as a GW source and as a possible g-mode excitation mechanism in 2D \citep{murphy:09, marek:09, muller:13}, and in 3D \citep{muller:12a}, although not established as a dominant source pre-explosion. Furthermore, \citet{torres-forne:19} (in 2D) point towards the stable inner core as the region responsible for the dominant emission in their study, and support the hypothesis that the dominant emission is always a g-mode \citep{torres-forne:19b}. For \citet{morozova:18} and \citet{sotani:2020} in 2D, and for \citet{radice:19} in 3D, their dominant source of emission after roughly 0.4\,s stems from the fundamental quadrupolar mode, having transitioned from a g-mode.

It is possible that the dominant GW emission is strongly model dependent and that the scenery of strong GW emission in CCSNe is as rich as the literature suggests. It is further possible that the situation is more generic and a stronger consensus will emerge as PNS asteroseismology techniques and classification methods become even more sophisticated and widely used. However, only a couple \citep{andresen:17, mezzacappa:20} spatially decompose their GW signals from the actual simulation. Thus, to shed light on the excitation mechanism, we encourage the combination of self-consistent perturbation analysis and spatially resolving the emission. 
In this work, we refrain from firmly diagnosing our dominant emission with a particular mode, we leave this to future work. Instead, we attempt to localize the source of the GWs in our simulations while simultaneously highlighting two analytical procedures, hoping to provide value to future studies in unveiling the mechanism and source of GWs:

(1) In Appendix\,\ref{Appendix:A} we remark that caution needs to be taken when calculating the spatial distribution of GWs since the often-used analytic expression for the first time derivative of the quadrupole moment (Eq.~\ref{eq:dIijdt} in Cartesian and Eq.~\ref{eq:dIzzdt} in spherical coordinates) neglects a surface term when considering only a portion of the PNS. This term, of course, vanishes when calculating the GW emission from the whole star. In that appendix we show how, when instead numerically taking two time derivatives of the quadrupole moment itself, the apparent emission region changes. In particular, emission that was previously seemingly from the convection zone, where the surface terms are largest, decreases.

(2) In Appendix\,\ref{Appendix:B} we highlight that the location where the energy density associated with a particular mode \citep{torres-forne:18, morozova:18, torres-forne:19, sotani:2020} is most concentrated, does not necessarily overlap with the region emitting the largest contribution of GWs due to the excitation of that mode. This is for two reasons, (a) because significantly more mass is located further out at larger radii, even though the energy density is lower and (b) the added lever arm of the radius (squared) in the quadrupole moment. As a consequence of this analysis, we show that the radial profile of the GW emission at any particular frequency, computed from either a perturbation analysis or dynamically from the simulation agree remarkably well in the PNS itself.

\begin{figure*}[ht]
\begin{center}
\includegraphics[width=\textwidth]{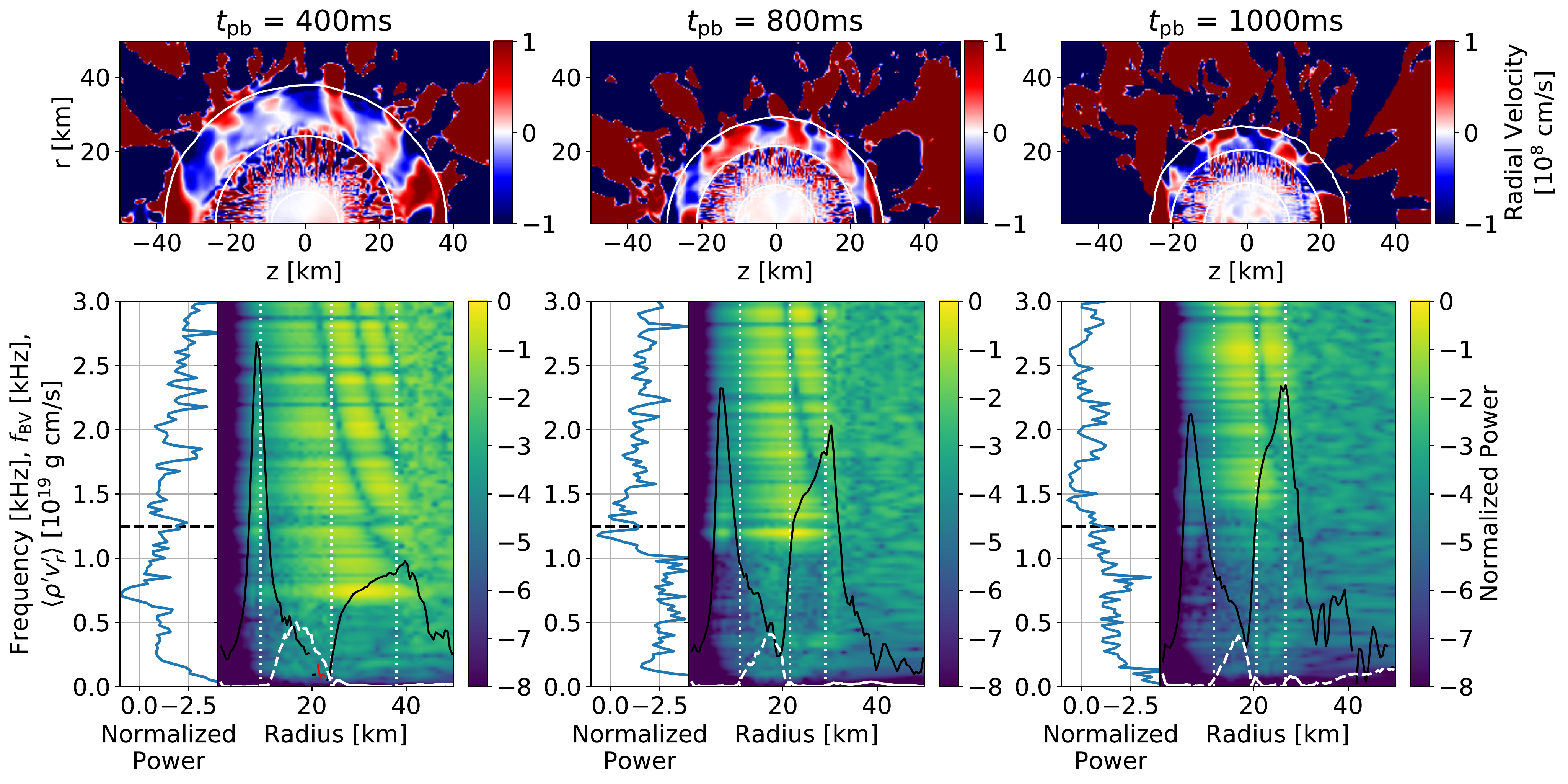}
\caption{Spatial distributions of the radial velocity (\textit{top panels}) where red (blue) colors denote positive (negative) radial velocities, as well as radial profiles of the GW spectrogram (\textit{bottom panels}) at times $t=400$\,ms (left), $t=800$\,ms (middle), and $t=1000$\,ms (right) after bounce for model \texttt{m0.75}. In contours (top) and vertical dashed lines (bottom) we denote the location of $\rho=10^{11}$, $10^{13}$, and $2\times10^{14}$\,g\,cm$^{-3}$.  These roughly divide the PNS core, PNS convection zone, the convectively-stable PNS surface layer, and the gain region (and beyond). To the left of each spectrogram profile we show the projected power spectral density, with an explicit dashed line at 1250\,Hz to highlight the location of the power gap.  On each of the spectrogram profiles we show the Brunt-V\"ais\"al\"a frequency (Eq.~\ref{eq:fbv}) in black (red denotes negative values of $f^2_\mathrm{BV}$) and the turbulent energy flux (Eq.~\ref{eq:fm}) in white (dashed denotes negative values).}
\label{fig:fvsr}
\end{center}
\end{figure*}

With these in mind, we attempt to localize the source of the GW emission in our 2D simulations and place the results in the context of the literature. In Figure~\ref{fig:fvsr}, we show the spatial distribution of the radial velocity (top panels) and radial profiles of the GW spectrogram (bottom panels) for $t=400$\,ms (left), $t=800$\,ms (middle), and $t=1000$\,ms (right) post-bounce for our baseline simulation, \texttt{m0.75}. The GW signal in these spectra are calculated in the post-processing of high cadence data by taking two numerical time derivatives of the contribution to the $zz$ component of the  quadrupole moment (i.e. Eq.~\ref{eq:quadrupole}) contained in each spherical shell (and multiplying with the constants outside the double derivative in Eq.~\ref{eq:radiation}). We note this is different than using a radially decomposed Eq.~\ref{eq:dIijdt} and taking one additional time derivative, we refer the reader to Appendix \ref{Appendix:A} for the important details.
We obtain the frequency dependence by multiplying this radially-decomposed GW signal with a 40\,ms Hann-window centered at each time specified above, and finally Fourier transform it. To the left of each spectrogram profile we show the normalized projected power spectral density. We note that both the profile and projection are logarithmic.  In each spectrogram profile we also plot the turbulent energy flux (white solid lines for positive values and dashed for negative values).  Similar to Eq.~$15-17$ in \cite{andresen:17}, the turbulent energy flux, $f_m$, is calculated via,
\begin{equation}
f_m = \langle \rho' v_r' \rangle\,,\label{eq:fm}
\end{equation}
\noindent
where angled brackets denote an angular average, $\rho$ denotes mass density, and $v_r$ is the radial velocity of the fluid. Here, the primed quantities are the deviations from the angle average, e.g. $\rho'~=~\rho - \langle \rho \rangle$. To reduce the stochastic noise of $f_m$ in our 2D simulations, the curves in Figure~\ref{fig:fvsr} are an average of all the turbulent energy flux profiles obtained over the 40\,ms window. We also include the Brunt-V\"ais\"al\"a frequency (black lines) following \cite{mezzacappa:20},
\begin{equation}
f^2_\mathrm{BV} = -\frac{\partial \phi/ \partial r}{(2\pi)^2 \rho}\left(\frac{\partial \rho}{\partial s}\bigg|_{P,Y_\mathrm{lep}} \frac{ds}{dr} +  \frac{\partial \rho}{\partial Y_\mathrm{lep}}\bigg|_{P,s} \frac{dY_\mathrm{lep}}{dr}\right)\,,\label{eq:fbv}
\end{equation}
where $Y_\mathrm{lep} = Y_e + Y_\nu$, and both $P$ and $s$ include contributions from the neutrinos as well.  For these neutrino contributions, we assume a thermal distribution of neutrinos at densities higher than $\sim \rho=10^{12}$\,g\,cm$^{-3}$ \citep{kaplan:14}.  While in principle the convective region should have negative $f^2_\mathrm{BV}$, we only see this for the outer part of the convection zone for the $t=400$\,ms profile, and therefore we rely on the turbulent energy flux to define the convection zone. For each of the three times, we mark the location of $\rho=10^{11}$, $10^{13}$, and $2\times10^{14}$\,g\,cm$^{-3}$ with dashed vertical lines in the spectrograms and as contours in the radial velocity plots.

\begin{figure}[ht]
\begin{center}
\includegraphics[width=\linewidth]{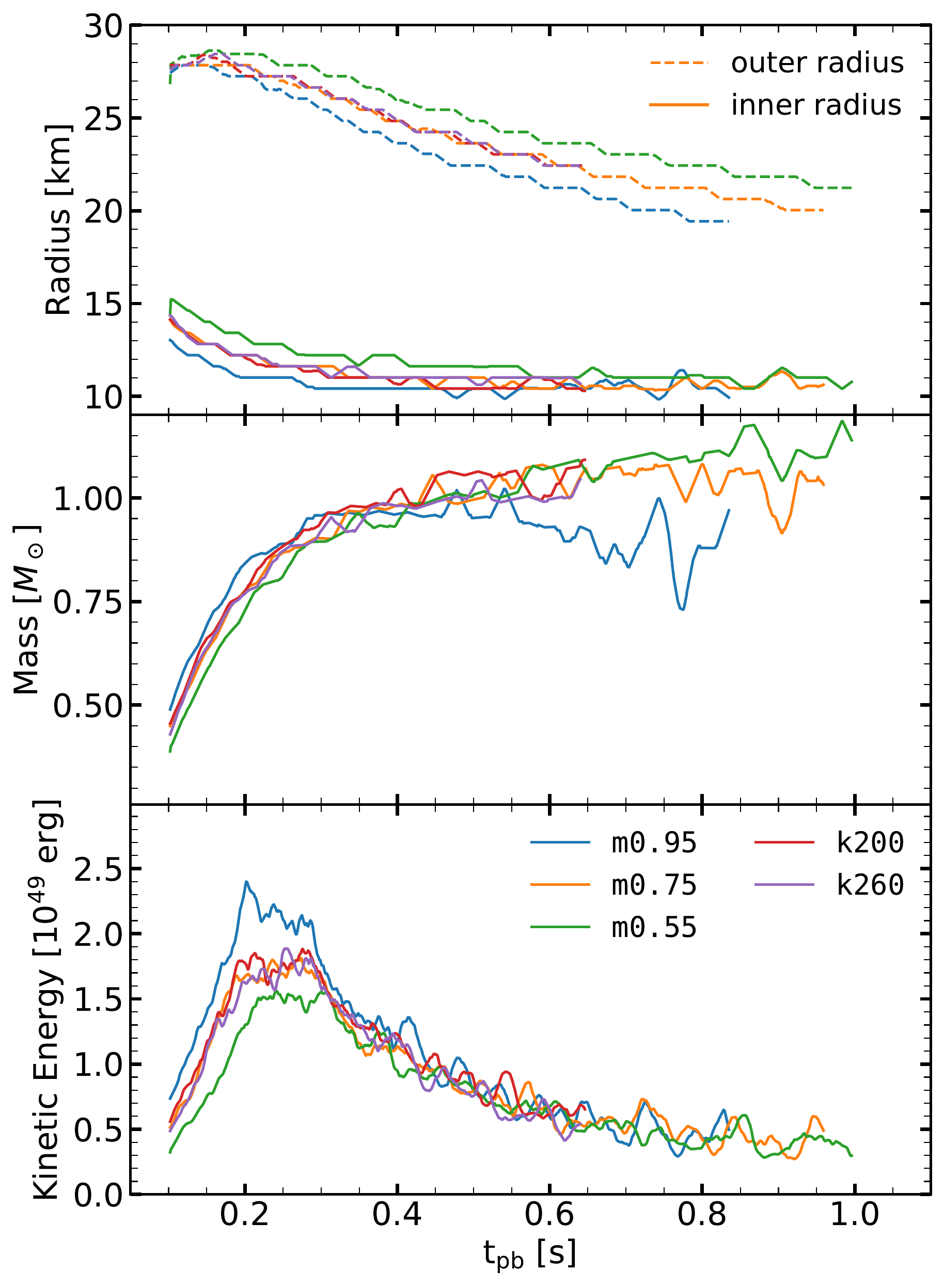}
\caption{\textit{Top panel}: Radial boundaries for the PNS convection zone as defined by the location where the turbulent flux is $f_m = 0.1 \times f^\mathrm{min}_m$  (inner) and $f_m = 0$ (outer). \textit{Middle panel}: PNS convection zone mass. \textit{Bottom panel}: PNS convection zone total kinetic energy. Data are smoothed over 20\,ms.}
\label{fig:pns_props}
\end{center}
\end{figure}

In Figure~\ref{fig:fvsr}, the convective region is traced very well by the width of the negative part of the turbulent energy flux profile. Inside the convection zone, heavier fluid is advected downwards while fluid that is lighter than average rises upwards. The turbulent energy flux will thus always be negative in the convective layer. The reverse is true in the overshoot layer above the PNS convection zone as the overshoot plumes are more dense than the local average, explaining the small positive peak in the turbulent energy flux just above the convection zone in Figure~\ref{fig:fvsr}. Hence, as seen in the figure, the radius corresponding to $\rho=2\times 10^{14}$, $10^{13}$, and $10^{11}$\,g\,cm$^{-3}$, roughly divides the PNS core, the PNS convection zone, the convectively-stable PNS surface layer, and the gain region (and beyond). This is further highlighted by the density contours in the radial velocity plots in the top panel of Figure~\ref{fig:fvsr} where funnels of fast-moving matter in opposite directions are seen between $\rho=10^{13}$\,g\,cm$^{-3}$ and $\rho=2\times 10^{14}$\,g\,cm$^{-3}$, namely convection sustained and driven by continued neutrino diffusion out of the core
and the maintenance of the lepton gradients. For completeness, we show in Figure~\ref{fig:pns_props} properties of the PNS convection zone over the course of the simulations for each non-rotating simulation.  Following the PNS radius trend in Figure~\ref{fig:spec_overlay_meff}, the PNS convection zone is located at a larger radius for simulations with a lower effective mass. While the smallest (largest) in volume, the \texttt{m0.95} (\texttt{m0.55}) simulation shows the largest (smallest) PNS convection zone mass and kinetic energy at early times ($t\lesssim 300$\,ms). At late times, the mass contained in the convection zone levels out and the ordering inverts, with the \texttt{m0.95} simulation having less mass than the others. This may be due to the explosion in the \texttt{m0.95} simulation. At the level explored here, $K_{\mathrm{sat}}$ does not impact any of the PNS convection properties.

Regarding the source of the GWs and the radially decomposed spectrograms in Figure~\ref{fig:fvsr}, resolving a precise location in the star where the emission occurs is difficult. The emission region is extended, which is supported by the results shown in Appendix~\ref{Appendix:B}, where we show, for these select times, the radial profile of the GW emission compared to the predictions from a perturbation analysis.   However, the figure provides a strong indication that the dominant emission in our simulations stems from the convectively stable surface layer, with a tendency towards the bottom end of this layer, the convective overshoot region. Characteristic of g-modes, the $\rho=10^{13}$\,g\,cm$^{-3}$ line, which closely traces the location of the overshoot region, crosses the Brunt-V\"ais\"al\"a frequency in close proximity to the dominant emission region, especially at $t=800$\,ms (middle), and $t=1000$\,ms (right). While our analysis suggests that the dominant source of GW emission does not stem from the convection zone itself, we cannot rule out PNS convection as a forcing mechanism that accounts for a substantial fraction of the GWs generated. In fact, we think this is plausible and supports the findings of \citet{andresen:17} who also find significant emission in the PNS convective overshoot region. On the other hand, considering the correlation we see between the GW amplitudes and the neutrino heating (\S \ref{sec:results:meff} and \S \ref{sec:results:rot}), as well as the non-correlation with the total PNS convection layer kinetic energy seen in Figure~\ref{fig:pns_props}, we neither rule out accretion plumes as a strong excitation mechanism. Furthermore, via entropy field plots we observe clear funnels of low-entropy matter (from the gain region) striking the PNS at times coinciding with some of the largest amplitude spikes in the GW signal. Perhaps these simulations serves as an example where both mechanisms are at work.\\

Several groups report on GW amplitudes decreasing by a factor $10-20$ when going from 2D to 3D \citep{andresen:17, oconnor-couch:18, mezzacappa:20}. This is generally expected since the axisymmetric assumption leads to an artificial enhancement of coherent motion. \citet{andresen:17} gather the contributions to the GW signal from different regions, such as the convection zone plus its overshoot layer (region A in their paper) and the PNS surface layer (region B in their paper). They report on a suppression of the GW signal by a factor of $\sim 10$ in region A when going from 2D to 3D, and a higher such factor in region B. \citet{andresen:17} discuss why a higher suppression of GWs excited by accretion plumes, compared to overshoot from below, is expected when going from 2D to 3D. In 2D, turbulence structures cascade in reverse \citep{xiao:09} and the Kelvin-Helmholtz instability is suppressed at the edge of supersonic downflows \citep{muller:15}, this may result in artificially high accumulation of the turbulent energy on large scales, and higher downflow velocities onto the PNS. Furthermore, \citet{andresen:17} demonstrate that in 2D, the frequency range of the chaotic turbulent downflows from the gain region may have a stronger overlap with the natural g-mode frequency, such that the accretion mechanism may produce more resonant excitation in 2D compared to 3D. Speculatively then, if both the convective overshoot and the accretion mechanisms drive the GWs in our simulations, and with these effects taken into account, 3D counterparts to our models may have convective overshoot as the main forcing of PNS oscillations, however, based in Appendix\,\ref{Appendix:B}, we do not expect the radial distributions to change dramatically.  \\

\subsection{Origin of the Power Gap}
\label{sec:results:gap}

\begin{figure}[ht]
\begin{center}
\includegraphics[width=\linewidth]{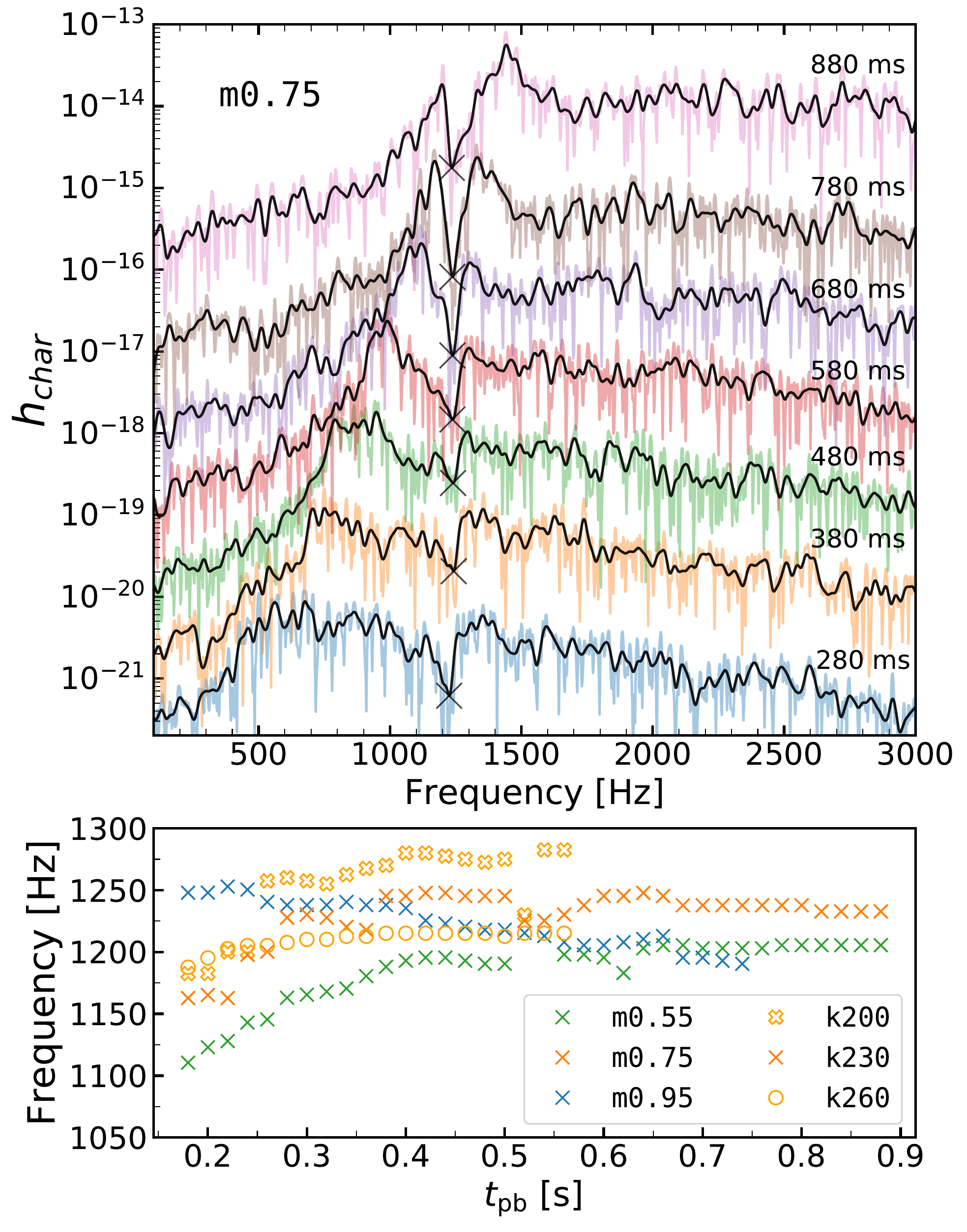}
\caption{\textit{Top panel}: Characteristic strain (at a 10\,kpc distance) calculated for seven times between 280$-$880\,ms post-bounce for the \texttt{m0.75} run (colored lines), with corresponding smoothed versions (black lines). We use a Hann function of 200\,ms centered at each time specified, windowing only that section of the GW signal (Figure~\ref{fig:hxd_meff}). Each but the first plot is shifted vertically using  logarithimically even multiplication factors. This figure provides yet another perspective of the drift of the peak frequency corresponding to the dominant mode frequency. We remark the large dip at $\sim 1.2$\,Hz and mark the local minimum of the smoothed functions which corresponds to the power gap. \textit{Bottom panel}: Applying the same technique to all non-rotating models, we track the frequency of the power gap over many time steps spaced 20\,ms apart. In the text we discuss the systematic ordering of the gap with the two EOS parameters and, in general, the central density.}
\label{fig:Hchar_seq}
\end{center}
\end{figure}

We end our results section by exploring the power gap present in all of our simulations.  We show the evolution of the characteristic strain for the \texttt{m0.75} run in the top panel of Figure~\ref{fig:Hchar_seq}, where we have sequentially windowed 200\,ms of the GW signal centered at the epoch we specify on the right hand side. For a direct comparison, each but the first strain is shifted vertically using even multiplication factors on a logarithmic scale. The strains are plotted in color, with smoothed versions in black. This figure provides yet another perspective of the drift of the peak frequency corresponding to the dominant PNS oscillation mode which goes from $\sim550$\,Hz at 280\,ms to $\sim1450$\,Hz at 880\,ms. The power gap, which we mark by crosses at the local minima of the black lines, is located around 1.2\,kHz and is particularly easy to see between $0.6-0.9$\,s when modes are excited both above and below the gap. In the bottom panel, we plot the location of the minimum over many time steps for the three models varying in effective mass and for the three models varying in incompressibility modulus. Recall that \texttt{m0.75} and \texttt{k230} are the same, and our baseline simulation. 

Addressing the models varying only in the effective mass first, before $\sim 0.3$\,s, the gap locations are ordered in frequency with the value of the effective mass. For \texttt{m0.55} and \texttt{m0.75}, which share the same general gap evolution, the frequency of the gap increases until it flattens out around $\sim 0.4$\,s and stays roughly constant thereafter. The gap evolution of the \texttt{m0.95} simulation differs from the others since its frequency decreases with time.

The evolution of the gap does not appear to change when varying the incompressibility modulus.  However the quantitative value of the gap frequency does.  It is located at lower frequencies for progressively higher values of $K_\mathrm{sat}$. Recall the {\tt k200} and {\tt k260} simulations are only carried out until $\sim 600$\,ms after bounce.

\citet{morozova:18} speculate that one may view the PNS as a coupled system of the inner core and the outer convectively stable shell, mediated by the PNS convection zone, such that the power gap may originate via a repelling interaction between a trapped PNS core mode and a surface layer mode. Their perturbation analysis roughly reproduces the morphology of the gap when plotting the frequency of modes after having moved the outer boundary to the core surface (their Figure~8). 

The different gap frequencies (bottom panel of Figure~\ref{fig:Hchar_seq}) in our study further hints towards an involvement of the inner core in producing the power gap. The incompressibility modulus has only a significant effect in the core, where the density is sufficiently high and the thermal component of the EOS is less dominant \citep{schneider:19}. Recall that the PNS radius defined at $\rho=10^{11}$\,g\,cm$^{-3}$ and the PNS convection zone properties, which reflects the structure between $\sim 10^{13}-10^{14}$\,g\,cm$^{-3}$, remain equal between the models varying in incompressibility modulus. Thus, the outer structure of the PNS alone is likely not the cause of the gap, nor sets the frequency. However this inner core structure may. Varying the incompressibility (effective mass) above and below baseline changes the central density by $\sim 4$\,\% ($\sim 12$\,\%). Furthermore, the central density and the gap frequency share the same ordering between all the models. That is, the higher the central density, the higher the frequency of the gap. The \texttt{m0.95} run is the only deviation to this trend due of its deviation in gap morphology.\\

Related to the gap origin suggested by \citet{morozova:18}, other asteroseismology studies see examples of avoided crossings between modes \citep{torres-forne:19, sotani:2020}. \citet{torres-forne:19} introduce mode grouping/classification via the shape of the mode functions, rather than by the number of nodes. With this method, some of their earlier-predicted avoided crossings (when counting nodes) are replaced by plain crossings, shedding light on the exchange of properties when modes cross. If what we see in our simulations is an avoided crossing, perhaps alike the analytical eigenmode solutions in \cite{sotani:2020}, the location of the avoided crossing coincides with the gap (Figure~\ref{fig:spec_meff}). While this is likely no coincidence, and would naturally explain the absence of emission in a small region in the spectrogram (where the modes interact), we have not untangled the process responsible for the power gap which persists during the entire simulation and postpone further investigation to future work. However, our preliminary perturbation analysis yields eigenmodes predicting the two main modes seen in the spectrograms, and further reveals that the \texttt{m0.95} model differs from the others by predicting another avoided crossing at an earlier epoch between less excited modes, a possible explanation for its unique gap morphology. From this preliminary analysis, we also see an eigenmode close to the gap whose energy is mostly confined in the inner core.

For completeness, we remark that the power gap is also evident (and marked with a black dashed line in the projected power spectral density) in the spatially decomposed spectrograms, Figures\,\ref{fig:fvsr}, \ref{fig:hpart_compare}. Other regions lack emission too, which very roughly correspond to nodes from p-modes above the gap and g-modes below (although these are not necessarily eigenmodes but perhaps sporadic wave-like transient perturbations). These ``g-mode'' nodes appear to converge at the gap, which in principle could produce it.

To finalize our discussion of the power gap we speculate on the reason why some investigations in the literature do not seem to show evidence for this gap \citep[e.g.][]{mezzacappa:20,torres-forne:19,jardine:21}. Our results, and those of \cite{morozova:18}, show clear evidence that the gap is tied to the physics at the highest densities in the inner ($\lesssim 10$\,km) core of the PNS. One potential cause for this disparity is the use of a spherical, 1D, inner core, which prevents aspherical motions, or in this case, potential mode interactions. \cite{morozova:18,oconnor-couch:18,radice:19} and this work, all see the power gap (or the power gap can be seen in the publicly available data) and do not have such an inner core. The one exception we can find is the work of \cite{pan:18}, who use a very similar setup to our FLASH simulations, but whose published GW spectrograms show no sign of a power gap.  On the other hand, several studies see a much more rich set of isolated excited modes \citep{torres-forne:19} compared to what is seen in our study.  It very well could be that aspects of our numerical setup are enhancing interactions between modes (including the interaction responsible for the power gap), further work still is needed to assess the power gap.

\section{Conclusions}
\label{sec:conclusions}
We have presented 9 axisymmetric simulations of CCSNe where we methodically vary experimentally accessible parameters in the EOS of dense nuclear matter. Our investigation focuses on the effective mass parameter $m^\star$ and the incompressibility parameter $K_{\mathrm{sat}}$, allowing values of $m^\star \in $ \{0.55, 0.75, 0.95\} and $K_{\mathrm{sat}} \in$ \{200, 230, 260\}\,MeV\,baryon$^{-1}$ which represent their estimated mean and 2$\sigma$ interval. For the values of $m^\star$ above and below baseline, the effect of a rotating progenitor is demonstrated using central angular speeds, $\Omega_0$, of both 1 and 2\,rad\,s$^{-1}$. 

In order to investigate the source of GWs and the mechanism which excites the oscillatory perturbations, we have spatially decomposed the GW emission, used linear perturbation analysis, extracted convection zone properties, and provided a thorough discussion of our results in the context of existing literature. First addressed by \citet{morozova:18}, we also report on the power gap and its EOS dependence.\\

We show that the dominant PNS oscillation mode is heavily dependent on the effective mass and attains a higher frequency when the effective mass parameter is increased. This is due to the more compact PNSs formed for EOSs with higher effective mass, as it affects the thermal pressure during PNS formation \citep{schneider:19}. As such, the neutrinosphere temperature increases with the effective mass, resulting in higher neutrino/anti-neutrino luminosities and mean energies, and thus, more neutrino heating. Aided by hydrodynamic instabilities like the SASI and convection, this increases the violent neutrino-driven instabilities in the gain region that drive convective overshoot via accretion-plumes striking the PNS. Hence, increasing the effective mass also increases the gravitational wave amplitude and the energy emitted in gravitational waves.

When including rotation, neutrino emission decreases and the gain region is less prone to instabilities. With central angular speeds of $\Omega_0 = 1$\,rad\,s$^{-1}$, the effective mass dependence of the dominant mode frequency is still clearly present, but the GW amplitudes are muted to similar heights. For $\Omega_0 = 2$\,rad\,s$^{-1}$, although different effective mass values yield different PNS radii, any clear differences in GWs are washed out by the centrifugally stabilized gain region. These simulations illustrate how rotation affects neutrino quantities and the PNS radii non-linearly. Further studies that allow central angular speeds to vary in value within $1-2$\,rad\,s$^{-1}$ are encouraged to find a critical value, above which the details of the dominant frequency mode becomes non-distinguishable to observations.

The simulations where $K_{\mathrm{sat}}$ varies show no significant differences in neither PNS compactness, PNS convection zone properties, neutrino emission, nor GW signatures. 
However, as changes in incompressibility can lead to very different \textit{cold}-NS mass-radius relationships, it may be that after a few seconds, once the star has radiated most of its binding energy away in neutrinos, some features in the GW spectra may depend on the isoscalar incompressibility $K_{\rm sat}$ or on the less-constrained isovector incompressibility $K_{\rm sym}$. 

For a distance of 10\,kpc, which is a relevant Galactic scale, these 2D CCSNe are detectable with the current generation of laser interferometers. However, recent CCSN studies in 3D highlight that such detections may be marginal for more realistic CCSNe \citep{oconnor-couch:18, mezzacappa:20, szczepanczyk:21}.

A spatial distribution of the GW emission shows that the dominant emission in our simulations stems from the convectively stable surface layer between densities of $\rho = 10^{11}$ and $\rho = 10^{13}$\,g\,cm$^{-3}$, with a tendency towards the convective overshoot region bordering the convection zone. The emission, which also extends into the convection zone, is in close agreement with a perturbation analysis (Appendix\,\ref{Appendix:B}) suggesting that the GW emission results from coherent quadrupolar oscillatory perturbations of the PNS.

We relay two viable excitation mechanisms of the perturbations (1) accretion plumes onto the PNS surface layer from the post-shock region with which we find a correlation in the GW amplitude via the neutrino heating that, together with the SASI, drives the turbulent convection. This mechanism is established in the literature \citep[e.g.][]{murphy:09}. And (2) overshoot into the surface layer from the convection zone below. The volume, mass, and kinetic energy inside the convection zone is effective mass dependent, but we find no apparent correlation between these properties and the GW amplitude. Since we see the signatures of overshoot via the turbulent energy flux, we can not rule out this mechanism as a viable forcing of the perturbations, especially in the light of the results of \citet{andresen:17} and in part \citet{mezzacappa:20}. 3D studies are required to establish the potency of this mechanism, as it may be washed out by artificially strong excitation from accretion plumes in 2D.

We highlight two analytical procedures that could provide value to future studies in unveiling the mechanism and source of GWs: In Appendix\,\ref{Appendix:A} we remark that caution needs to be taken when calculating the spatial distribution of GWs utilizing the analytic expression for the first time derivative of the quadrupole moment (Eq.~\ref{eq:dIijdt} in Cartesian and Eq.~\ref{eq:dIzzdt} in spherical coordinates). In Appendix\,\ref{Appendix:B} we highlight that the location where the energy density associated with a particular mode \citep{torres-forne:18, morozova:18, torres-forne:19, sotani:2020} is most concentrated, does not necessarily overlap with the region emitting the largest contribution of GWs due to the excitation of that mode.

Related to the origin of GWs is the presence of a narrow ($\Delta f \sim$ 50\,Hz), high frequency ($\sim 1.2$\,kHz), and long lasting region in the spectrograms that lack emission, the ``power gap". We find that the frequency of the power gap changes when varying both the effective mass and the incompressibility modulus, and is generally ordered by the value of the central density. This, combined with more detailed arguments (\S\ref{sec:results:gap}) and previous analysis of \citet{morozova:18}, strongly hints towards an involvement of the inner core in producing the gap. The modes in the spectrogram appear to undergo an avoided crossing at a frequency coinciding with that of the gap. This is supported by our preliminary perturbation analysis which further indicates that the mode closest to the gap has its energy mostly confined to the inner core.

\begin{acknowledgments}
We thank Viktoriya Morozova and Michael Pajkos for in-depth discussions during the development of this work. This work is supported by the Swedish Research Council (Project No. 2020-00452). The simulations were enabled by resources provided by the Swedish National Infrastructure for Computing (SNIC) at PDC and NSC partially funded by the Swedish Research Council through grant agreement No. 2016-07213.
\end{acknowledgments}

\software{FLASH \citep{fryxell:00}, NuLib \citep{oconnor:15}, Matplotlib \citep{hunter:07}, NumPy \citep{harris2020array}, SciPy \citep{2020SciPy-NMeth}, yt \citep{turk:11}}

\appendix

\section{Spatial distribution of gravitational-wave emission}\label{Appendix:A}
In this Appendix, we give the formulae to acquire the spatial distribution of gravitational-wave (GW) emission. Below we use the spherical coordinate system $(r,\theta,\phi)$. In axisymmetry, the mass quadrupole moment of a shell centering at $r$ with a thickness of $\Delta r$ is
\begin{equation}\label{eq:part_Izz}
    I_{zz} = \frac{2}{3} \int_{r-\Delta r/2}^{r+\Delta r/2} \rho r'^2 P_2(\cos\theta) dV,
\end{equation}
where $P_2$ is the Legendre polynomial of degree 2, and $dV=r'^2\sin\theta dr' d\theta d\phi$. According to Eq.~\ref{eq:radiation}, the GW strain contributed by this shell is
\begin{equation}\label{eq:part_hplus}
    h_+(r,t)=\frac{2}{D}\frac{G}{c^4} \frac{d^2}{dt^2} I_{zz}.
\end{equation}
In the main text, we perform numerical differentiation of $I_{zz}$ with respect to $t$ twice to calculate shell contribution $h_+(r,t)$ with $\Delta r=1$~km. The Fourier transform of $h_+(r,t)$ within a temporal window of 40~ms gives $\tilde{h}_+(r,f)$. In Figure~\ref{fig:fvsr} we plot the normalized power $|\tilde{h}_+(r,f)|^2$ as radial profiles of the GW spectrogram.

\begin{figure*}[t]
    \centering
    \includegraphics[width=0.97\textwidth]{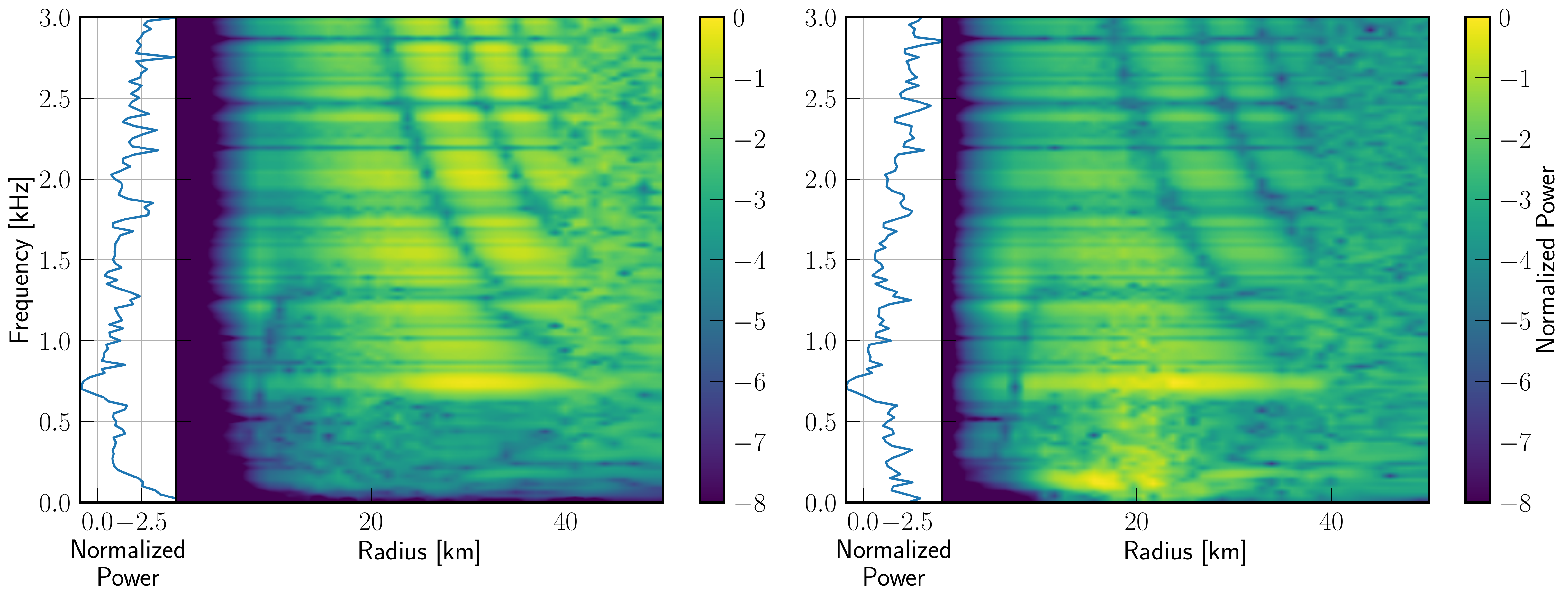}
    \caption{Radial profiles of GW spectrograms at $t=$0.4~s after bounce calculated by numerical differentiation of $I_{zz}$ (left panel) and $dI_{zz}/dt$ without the surface term (right panel).}
    \label{fig:hpart_compare}
\end{figure*}

Generally, one wants to avoid numerical differentiation to minimize numerical noise \citep{finn-evans:90}. With the mass conservation equation $\frac{\partial\rho}{\partial t}+\nabla\cdot (\rho \vec{v})=0$, one can avoid the first numerical differentiation of $I_{zz}$ as
\begin{equation}
\frac{dI_{zz}}{dt} =  \frac{2}{3} \int_{r-\Delta r/2}^{r+\Delta r/2} \frac{\partial\rho}{\partial t} r'^2 P_2(\cos\theta) dV = -\frac{2}{3} \int_{r-\Delta r/2}^{r+\Delta r/2} \nabla\cdot (\rho \vec{v}) r'^2 P_2(\cos\theta) dV.
\end{equation}
Partial integration is used to further avoid the spatial differentiation. For example, the radial derivative can be avoided by
\begin{align}
    \Big(\frac{dI_{zz}}{dt}\Big)_r & = -\frac{2}{3} \int_0^{2\pi}d\phi \int_0^{\pi}P_2(\cos\theta)\sin\theta d\theta \int_{r-\Delta r/2}^{r+\Delta r/2} \frac{\partial}{\partial r'}(r'^2 \rho v_r) r'^2 dr' \\
    & = -\frac{2}{3} \int_0^{2\pi}d\phi \int_0^{\pi}P_2(\cos\theta)\sin\theta d\theta \Big( r'^4\rho v_r|_{r-\Delta r/2}^{r+\Delta r/2} -2\int_{r-\Delta r/2}^{r+\Delta r/2} r'^3\rho v_r dr'\Big).\label{eq:dIzzdt}
\end{align}
Note that there is a surface term $r'^4\rho v_r|_{r-\Delta r/2}^{r+\Delta r/2}$ from the partial integration. This term can be ignored when calculating $dI_{zz}/dt$ of the whole star, because it vanishes at $r=0$ and at the stellar surface where $\rho=0$. However, the surface term cannot be dropped for an arbitrary mass shell. In Figure~\ref{fig:hpart_compare} we compare the spatial GW spectrogram evaluated by numerical differentiation of $I_{zz}$ (left panel) and $dI_{zz}/dt$ (right panel) calculated as 
\begin{equation} \label{eq:dIdt}
    \frac{dI_{zz}}{dt}=\frac{2}{3} \int_{r-\Delta r/2}^{r+\Delta r/2} \rho r' \Bigg(2v_r P_2(\cos\theta) + v_\theta \frac{\partial P_2(\cos\theta)}{\partial \theta} \Bigg) dV,
\end{equation} 
where the surface terms are ignored. Though the total GW emission is not affected due to canceling of the surface terms (except for the noise due to numerical differentiation), the detailed spatial distribution is clearly different. This comparison shows that using Eq.~\ref{eq:dIdt} can lead to a problematic results in the spatial contribution to GW emission, especially if $\Delta r$ is small.

\section{Comparison between simulation and perturbation analysis}\label{Appendix:B}
We follow \cite{ryan:20} to perform a perturbation analysis which is consistent with our pseudo-Newtonian hydrodynamic simulations. Eqs. 3.12-3.15 in \cite{ryan:20} are solved with a 4-stage Runge-Kutta method to get the quadrupolar ($l=2$) perturbative mode functions, namely, the radial displacement $\eta_r$, tangential displacement $\eta_\bot$, and density $\delta \hat{\rho}$ as a function of radii. Here, we do not fix the outer boundary condition to get the eigenmodes, but set the mode frequency in the perturbative equations to be the peak GW frequency and acquire the corresponding mode functions for the background fluid at a specific time. The perturbative quadrupole moment responsible for GW emission is
\begin{equation}
    \delta Q = \int r^4 \delta \hat{\rho} dr. \label{eq:perturbQ}
\end{equation}
In the top panel of Figure~\ref{fig:perturb} we compare the radial profiles of GW emission ($\tilde{h}_+(r,f)$ in Appendix~\ref{Appendix:A}) in the simulations with $C r^4 \delta \hat{\rho}$ from the perturbation analysis, where a multiplication constant $C$ is used to match the maximums between them. Agreement is found inside the PNS ($\rho \ge 10^{11}$~g~cm$^{-3}$), suggesting that GW emission results from the coherent quadrupolar oscillations of the entire PNS. In the bottom panel of Figure~\ref{fig:perturb} we plot the radial profiles of the kinetic energy density of the perturbative mode as \citep[Eq. 59 in][]{torres-forne:19}
\begin{equation} \label{eq:ekin}
    E_{\rm kin} = \rho \Big[\eta_r^2+l(l+1)\Big(\frac{\eta_\bot}{r}\Big)^2\Big],
\end{equation}
where $l=2$. Similar to \cite{torres-forne:19,morozova:18}, the kinetic energy density has its maximum inside the inner core ($\sim10$ km). However, the kinetic energy which has a $r^2$ from the volume $dV$ resembles the distribution of GW emission (except for the leakage to outside PNS at $t_{\rm pb}=800$~ms). We assert from this that using the energy density alone to quantify the location of GW emission is inappropriate. Details on the perturbation analysis such as eigenmodes analysis is still under investigation and will be described elsewhere. 

\begin{figure}
    \centering
    \includegraphics[width=0.97\textwidth]{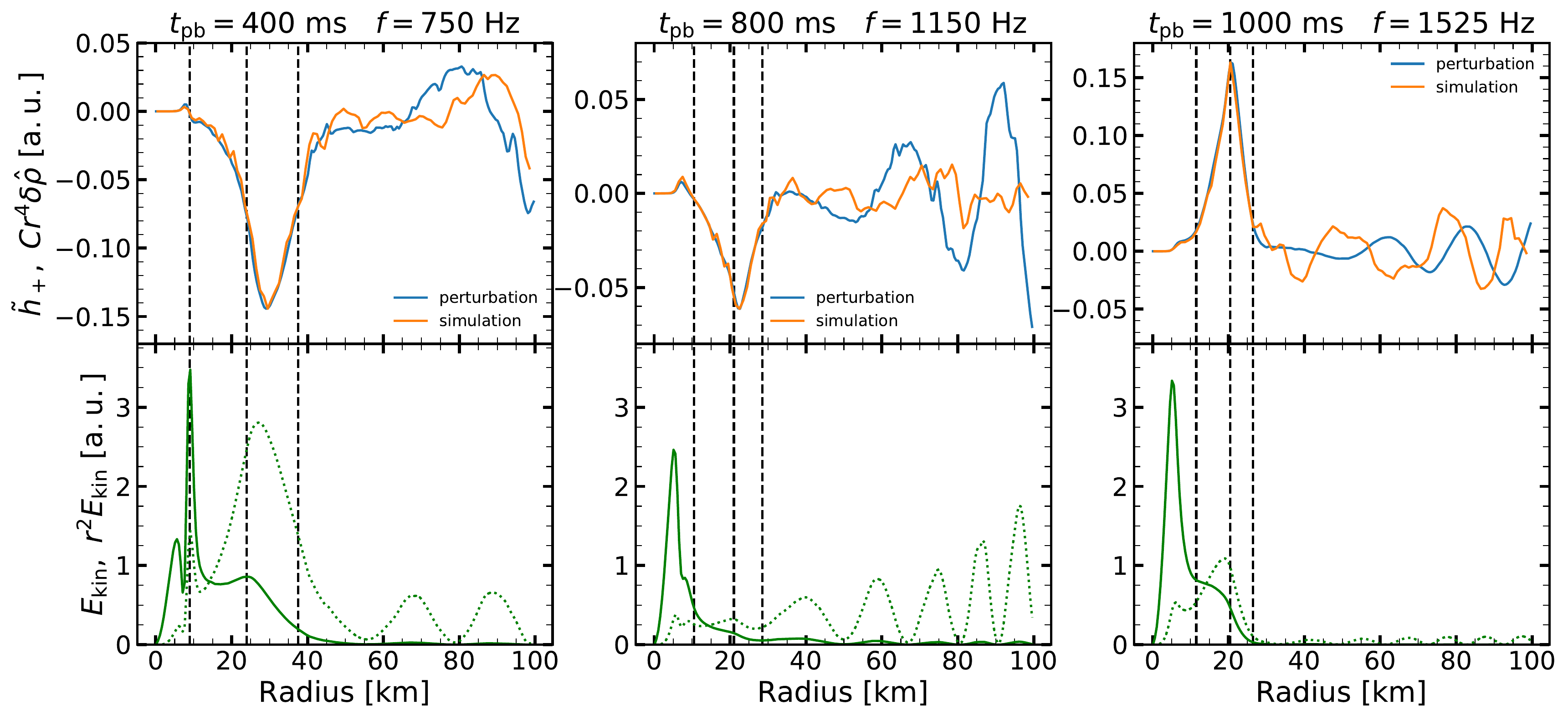}
    
    \caption{Comparison between perturbation analysis and simulations with $m^{\star}=0.75$, for GW peak frequencies at $t=400,~800,~1000$~ms after bounce, from left to right. \textit{Top panel}: simulation is $\tilde{h}_+(r,f)$ described in Appendix~\ref{Appendix:A} and perturbation is $Cr^4\delta\hat{\rho}$ in Eq.~\ref{eq:perturbQ}. \textit{Bottom panel}: solid line is $E_{\rm kin}$ in Eq.~\ref{eq:ekin} and dashed line is scaled $r^2E_{\rm kin}$. Vertical black dashed lines in each panel are locations of $\rho=2\times10^{14}, ~10^{13}, ~10^{11}$~g~cm$^{-3}$, from left to right.}
    \label{fig:perturb}
\end{figure}

\end{document}